\documentclass[preprint,12pt]{elsarticle}

\usepackage{graphicx}
\usepackage{amssymb}

\usepackage{amsmath}

\DeclareMathAlphabet{\mathpzc}{OT1}{pzc}{m}{it}

\biboptions{sort}

\journal{arXiv}

\usepackage[section]{placeins}

\usepackage{url}

\usepackage{framed}
\usepackage{multicol}
\usepackage{nomencl}
\makenomenclature
\renewcommand*\nompreamble{\begin{multicols}{2}}
\renewcommand*\nompostamble{\end{multicols}}

\begin{document}

\begin{frontmatter}


\title{Mixed finite elements for \\
convection-coupled phase-change in enthalpy form: \\
Open software verified and applied to 2D benchmarks}


\author[Aachen]{Alexander G. Zimmerman}
\ead{zimmerman@aices.rwth-aachen.de}

\author[Aachen,Göttingen]{Julia Kowalski}

\address[Aachen]{RWTH Aachen Univ., AICES, Schinkelstraße 2, 52062 Aachen, Germany}
\address[Göttingen]{Univ. of Göttingen, Comput. Geosci., Goldschmidtstr. 1, 37077 Göttingen, Germany}

\begin{abstract}

Melting and solidification processes are often affected by natural convection of the liquid, posing a multi-physics problem involving fluid flow, convective and diffusive heat transfer, and phase-change reactions.
Enthalpy methods formulate this convection-coupled phase-change problem on a single computational domain.
The governing equations can be solved accurately with a monolithic approach using mixed finite elements and Newton's method.
Previously, the monolithic approach has relied on adaptive mesh refinement to regularize local nonlinearities at phase interfaces.
This contribution instead separates mesh refinement from nonlinear problem regularization and provides a continuation procedure which robustly obtains accurate solutions on the tested 2D uniform meshes.
A flexible and extensible open source implementation is provided.
The code is formally verified to accurately solve the governing equations in time and in 2D space, and convergence rates are shown.
Two benchmark simulations are presented in detail with comparison to experimental data sets and corresponding results from the literature, one for the melting of octadecane and another for the freezing of water.
Sensitivities to key numerical parameters are presented.
For the case of freezing water, effective reduction of numerical errors from these key parameters is successfully demonstrated.
Two more simulations are briefly presented,
one for melting at a higher Rayleigh number and one for melting gallium.

\end{abstract}


\end{frontmatter}

\begin{table*}[!t]

   \begin{framed}
   
     \printnomenclature
     
   \end{framed}
   
\end{table*}


\nomenclature[1\(t\)]{$t$}{Time}

\nomenclature[1\(x\)]{$\mathbf{x}$}{Position}

\nomenclature[1\(x\)]{$x$}{Horizontal position}

\nomenclature[1\(y\)]{$y$}{Vertical position}


\nomenclature[2\(p\)]{$p$}{Pressure}

\nomenclature[2\(u\)]{$\mathbf{u}$}{Velocity}

\nomenclature[2\(T\)]{$T$}{Temperature}

\nomenclature[3\(nu\)]{$\nu$}{Kinematic viscosity}

\nomenclature[3\(mu\)]{$\mu$}{Dynamic viscosity}

\nomenclature[3\(rho\)]{$\rho$}{Density}

\nomenclature[3\(C\)]{$C$}{Volumetric heat capacity}

\nomenclature[3\(kappa\)]{$\kappa$}{Thermal conductivity}

\nomenclature[3\(f\)]{$f$}{Volume fraction}

\nomenclature[3\(phi\)]{$\phi$}{Regularized volume fraction}

\nomenclature[3\(T_m\)]{$T_m$}{Phase-change temperature}

\nomenclature[4\(l\)]{$l$}{Subscript for liquid}

\nomenclature[4\(s\)]{$s$}{Subscript for solid}


\nomenclature[5\(Pr\)]{$\mathrm{Pr}$}{Prandtl number}

\nomenclature[5\(Ra\)]{$\mathrm{Ra}$}{Rayleigh number}

\nomenclature[5\(Re\)]{$\mathrm{Re}$}{Reynolds number}

\nomenclature[5\(Ste\)]{$\mathrm{Ste}$}{Stefan number}

\nomenclature[5\(X\)]{$\mathrm{X}$}{Length scale}

\nomenclature[5\(U\)]{$\mathrm{U}$}{Speed scale}

\nomenclature[5\(delta T\)]{$\delta \mathrm{T}$}{Temperature scale}

\nomenclature[6\(b\)]{$b$}{Buoyancy function}

\nomenclature[6\(F\)]{$\mathcal{F}$}{Nonlinear problem residual}

\nomenclature[6\(w\)]{$\mathbf{w}$}{System solution}


\nomenclature[7\(h\)]{$h$}{Mesh cell size}

\nomenclature[7\(Delta t\)]{$\Delta t$}{Time step size}

\nomenclature[7\(sigma\)]{$\sigma$}{Regularization parameter}

\nomenclature[7\(tau\)]{$\tau$}{Velocity relaxation factor}

\nomenclature[7\(q\)]{$q$}{Quadrature degree}


\nomenclature[8\(n\)]{$n$}{Discrete time index}

\nomenclature[8\(j\)]{$j$}{Newton iteration index}

\section{Introduction}

Melting and solidification processes in phase-change materials are relevant to many areas of engineering and scientific research.
Examples include the casting of metals, storage of solar and thermal energy as latent heat, and modeling phase-change processes in the cryosphere, e.g. as relevant for climate projections.
In some regimes, the phase-change process is substantially affected by convection in the liquid phase, which has been demonstrated in a number of experiments \cite{sparrow1978experiments, sparrow1979freezing, okada1984analysis, kowalewski1999freezing, schuller2017integrated}.

Any spatiotemporally resolved simulation of this multi-physics process is a mathematical and computational challenge.
Common modeling strategies couple systems of multi-parameter nonlinear partial differential equations resulting from the balance laws of mass, momentum, and energy. 
For realistic scenarios, solving the system requires efficient and robust numerical methods.
The design and validation of models for different regimes and the formulation of accurate solution methods are areas of ongoing research.

The current work focuses on pure materials at macroscopic scale, for which melting and freezing occur at one specific temperature, namely the melting/freezing point $T_m$. 
The corresponding phase-change process is commonly referred to as \textit{isothermal} phase-change \cite{voller1987enthalpy}. 
Isothermal phase-change produces well defined, distinct phase interfaces at small spatial scales. 
There are many other materials that do not fall into this class, e.g. alloys or sea ice, and that produce substantial mushy regions, i.e. regions containing liquid and solid phases.
For such other materials, melting and freezing processes are bounded by solidus and liquidus temperatures which can both vary depending on the local material composition.
Isothermal phase-change can hence be interpreted as the limiting case, in which the solidus and liquidus temperatures coincide and are constant.

Many computational fluid dynamics (CFD) applications are concerned with a physical setting that consists of entirely gas or entirely liquid materials, sometimes extending into multi-phase gas-liquid applications.
In these cases, solids exist only as boundaries and are usually fixed in space and time.
Simulating convection-coupled melting and freezing requires the extension of CFD into multi-phase liquid-solid applications.
This involves modeling the spatiotemporally evolving interfaces between liquids and solids.
For this, one can use either an interface tracking or an interface capturing method \cite{voller1987enthalpy, alexiades1993mathematical}.
Interface tracking methods explicitly track the phase interfaces, solve separate systems of governing equations in each phase domain, and enforce coupling constraints at the interfaces.
Interface capturing methods instead solve one system of governing equations on a single domain that is occupied by both phases.
Interface capturing handles phase interfaces implicitly.
The phase interfaces can be post-processed as an explicit function of the solution.

The current work uses an interface capturing method that relies on an enthalpy formulation of the phase-change process. 
This is often referred to as an enthalpy method \cite{voller1987enthalpy}.
Generally, enthalpy methods write the energy balance in terms of both the temperature and the enthalpy,
and hence must be closed with an equation relating these two quantities.
For incompressible materials, this relation is well-understood and can be easily phrased in terms of the liquid and solid phase volume fractions. 
For isothermal phase-change processes with liquid and solid regions respectively above and below the melting/freezing temperature, the phase can be written as a function of only the temperature.
In these cases, the enthalpy can be eliminated from the system as an unknown.
This choice results in an energy balance taking the form of a standard convection-diffusion equation extended by a source term which accounts for the gain or loss of latent heat at the phase interface. 
Enthalpy methods have been applied extensively to convection-coupled phase-change \cite{voller1987enthalpy, voller1987fixed, brent1988enthalpy, alexiades1993mathematical, giangi2000natural, evans2007temporal, belhamadia2012enhanced, danaila2014newton, zimmerman2017monolithic, rakotondrandisa2019numerical, woodfield2019stability},
but there remain many opportunities for improvement,
e.g. regarding their dependency on numerical parameters and computational feasibility in three spatial dimensions. 

In the context of enthalpy methods, a variety of techniques have been applied to modeling the velocity in the solid phase.
These are sometimes referred to as a \textit{solid velocity correction} \cite{wang2010comprehensive}.
Different approaches are compared in \cite{voller1987enthalpy, voller1990fixed}.
Three primary approaches to solid velocity correction have prevailed: 
\begin{enumerate}
    \itemsep0em
    \item modifying the algorithm to set the velocity to zero in the solid \cite{wang2010comprehensive},
    \item adding a phase-dependent source term to the momentum equation, forcing the velocity to relax to zero in the solid \cite{belhamadia2012enhanced, schuller2017integrated, rakotondrandisa2019numerical}, and
    \item prescribing a phase-dependent viscosity with orders of magnitude larger values in the solid than in the liquid \cite{danaila2014newton, zimmerman2017monolithic}.    
\end{enumerate}
Of these, only the third approach can physically model solid regions which are not attached to stationary boundaries.
The first approach provides a \emph{brute force} solution and allows no velocity dampening close to the phase interface. 
The second and third approaches both rely on additional numerical parameters that can affect the evolution of phase interfaces. 
Critically, these additional parameters must be calibrated based on experimental data, as seen in \cite{danaila2014newton, rakotondrandisa2019numerical, rakotondrandisa2020toolbox}.
This need for calibration may undermine the predictiveness of the models, because it is not clear to what extent the calibrated parameters also apply to other physical regimes or geometric settings.

For any chosen approach to solid velocity correction, the abrupt change in velocity at phase interfaces disrupts nonlinear solver convergence \cite{brent1988enthalpy}.
A common regularization approach is to introduce a temperature range in which the phase-change occurs rather than restricting it to exactly the melting/freezing point.
There seems to be a strong analogy between this type of numerical method and the physical situation of non-isothermal phase-change for non-pure materials.
In both situations, the community tends to refer to the lower and upper temperature bounds respectively as solidus and liquidus temperatures.
It is, however, important to distinguish cases in which solidus and liquidus temperatures are physically motivated versus cases in which they are introduced as numerical parameters to stabilize an otherwise unstable computational model.
The current work includes sensitivity studies where the mesh size is held constant and the phase interface regularization is reduced until it does not substantially affect the size of the artificial mushy region.
The minimum extent of the artificial mushy region will always be constrained by the size of the mesh cells.

Enthalpy methods define a system of partial differential equations which must be discretized in space and time.
Many approaches have been applied for spatial discretization, including finite differences \cite{voller1987enthalpy, voller1987fixed, brent1988enthalpy}, finite volumes \cite{giangi2000natural, evans2007temporal, wang2010comprehensive, schuller2017integrated}, the finite element method (FEM) with operator splitting \cite{belhamadia2012enhanced}, and monolithic mixed FEM \cite{danaila2014newton, zimmerman2017monolithic, rakotondrandisa2019numerical, belhamadia2019adaptive, woodfield2019stability, alvarez2019newmixed, rakotondrandisa2020toolbox}.
Mixed FEM methods come both in purely primal forms \cite{danaila2014newton, zimmerman2017monolithic, rakotondrandisa2019numerical, belhamadia2019adaptive, woodfield2019stability, rakotondrandisa2020toolbox} and also in forms with auxiliary variables \cite{woodfield2019stability, alvarez2019newmixed}.
The cited primal formulations all used Taylor-Hood elements for the pressure-velocity system.
Of these, \cite{belhamadia2019adaptive, woodfield2019stability, rakotondrandisa2020toolbox} used quadratic elements for the temperature, making the spatial discretization method theoretically second order accurate for all of the solution fields.
Only recently has formal analysis of the stability, and numerical accuracy verification, of mixed FEM schemes for enthalpy formulated convection-coupled phase-change begun \cite{woodfield2019stability, alvarez2019newmixed}.
Both articles use the method of manufactured solutions to verify the spatial order of accuracy.
Temporal accuracy is also verified using manufactured solutions in \cite{woodfield2019stability}.
Typically, temporal discretization is performed with finite differences.
Second-order accurate fully implicit finite difference temporal discretization was advocated in \cite{evans2007temporal} and subsequently used in \cite{belhamadia2012enhanced, rakotondrandisa2019numerical, woodfield2019stability, rakotondrandisa2020toolbox}.

Based on the primal mixed finite element approach in \cite{danaila2014newton}, an open source implementation was published in \cite{zimmerman2017monolithic}.
Both \cite{danaila2014newton} and \cite{zimmerman2017monolithic} used adaptive mesh refinement (AMR), albeit with alternative formulations.
While results in \cite{zimmerman2017monolithic} were promising, it was later difficult to apply its presented implementation to solidification problems.
Furthermore, while that implementation could adaptively refine meshes, adaptive coarsening was not implemented.
The lack of mesh coarsening capability was computationally demanding for simulations run over a long simulated time with the phase interface moving large distances through the domain.

There are some drawbacks to any AMR approach.
For example, the benefit of reduced degrees of freedom competes with the cost of dynamic re-meshing, especially given that the phase interface moves throughout the spatial domain in time.
Furthermore, the benefit of per-cell accuracy from locally refined meshes competes with accumulating interpolation errors which come from transferring solutions between non-matching meshes of different time steps.
Still, application of AMR methods to convection-coupled phase-change is a promising, successful, and active (e.g. \cite{rakotondrandisa2019numerical, belhamadia2019adaptive, rakotondrandisa2020toolbox}) field of research,.

Another key limitation, noted in \cite{zimmerman2017monolithic}, is that AMR requires an \textit{a posteriori} error estimate, which in turn requires first obtaining a solution on a coarse or manually refined mesh.
Convergence of the nonlinear solution methods presented in \cite{danaila2014newton} and \cite{zimmerman2017monolithic} both required careful preparation of the problem setting, in particular relying on manual local refinement of the initial mesh, carefully chosen time step sizes, and artificial initialization of the new phase in one-way melting or solidification processes.
The current paper sets aside AMR to focus on more reliably obtaining solutions on uniform grids, including relatively coarse grids which could be used, for example, to automatically initialize AMR methods.

This article is structured as follows:
Section \ref{section:GoverningEquations} presents an enthalpy formulation for convection-coupled phase-change processes with a generalized approach to phase-interface regularization.
Section \ref{section:NumericalMethods} presents a monolithic solver approach with mixed finite elements, implicit time discretization, and Newton's method.
Motivated by \cite{brent1988enthalpy, zimmerman2017monolithic}, a continuation procedure is introduced with the goal of robustly solving the nonlinear problem without local mesh refinement, which is made difficult by the irregularity at the phase interface.
Section \ref{section:ImplementationAndVerification} introduces a new open-source implementation and describes the open-source libraries on which it depends.
Furthermore, empirical convergence rates are shown for the spatial and temporal discretizations using the method of manufactured solutions.
Section \ref{section:Validation} presents benchmark simulation results for the melting of octadecane and freezing of water, both on unit square geometries.
For both, sensitivities to numerical parameters are reported and the resulting time evolutions of the phase interface are compared to experimental data from the literature.
Furthermore, brief comparisons are made to the simulation results in \cite{rakotondrandisa2020toolbox}.
Section \ref{section:Examples} briefly presents two additional example simulations on rectangular geometries,
one for melting octadecane at a higher Rayleigh number and one for melting gallium.

\section{Physical and mathematical model} 
\label{section:GoverningEquations}

\subsection{Physical regime}

The current work focuses on simulating melting and solidification processes that are strongly influenced by convection of the liquid phase.
Particularly, the focus is on regimes where there is a strong two-way coupling between the fluid dynamics and the evolving shapes of phase interfaces.
It is assumed that all solid regions are attached to stationary boundaries.
The liquid is assumed as an incompressible Newtonian fluid undergoing natural convection.
Following a Boussinesq assumption, the fluid density is assumed to only depend on the temperature.
In this case, the flow is driven by buoyancy resulting from temperature gradients.
Furthermore, the focus is on pure materials such as octadecane and distilled water.
Such materials undergo isothermal phase-change \cite{voller1987enthalpy, brent1988enthalpy}.
This means that melting and solidification occur at a single temperature.

\subsection{Phase interface regularization}

Consider a bulk phase-change material which can either be fully liquid, fully solid, or have both liquid and solid regions.
Any small reference volume of this material has a liquid volume fraction $f_l$ and a solid volume fraction $f_s$.
Assuming there are no other phases, e.g. no gas inclusions, $f_l + f_s = 1$.
When the material is fully liquid, $f_l = 1$. When it is fully solid, $f_l = 0$.
In this case, the value of $f_l$ can be used as a substitute for the phase state.

Isothermal phase-change processes are characterized by having distinct interfaces between phases.
At these interfaces, the temperature of the material is at the melting/freezing point $T = T_m$.
In liquid regions, $T > T_m$.
In solid regions, $T < T_m$.
This means that the liquid volume fraction $f_l$ can be reconstituted from the temperature alone, i.e.
\begin{equation} \label{eq:LiquidVolumeFraction}
    f_l = 
    \begin{cases}
        0, & T < T_m \\
        1, & T > T_m
    \end{cases}
\end{equation}
With this model, phase interfaces are theoretically infinitesimally thin lines in two-dimensional space or surfaces in three-dimensional space.
The phase interfaces can be regularized by convoluting $f_l$ with a Gaussian kernel, e.g.
\begin{equation} \label{eq:Gaussian}
    \zeta(T) = \frac{1}{\sigma \sqrt{2 \pi}} \mathrm{exp}\left(\frac{-T^2}{2 \sigma^2}\right)
\end{equation}
yielding 
\begin{equation} \label{eq:RegularizedLiquidVolumeFraction}
    \phi_l = (f_l * \zeta)(T) = \frac{1}{2}\left(1 + \mathrm{erf}\left(\frac{T - T_m}{\sigma\sqrt{2}}\right) \right)
\end{equation}
This introduces a parameter, $\sigma$, which is the standard deviation of the Gaussian and is always positive.
As $\sigma$ approaches zero, $\phi_l$ approaches $f_l$.
With too large of $\sigma$, substantial regions may develop where $0 < \phi_l < 1$. 
In the context of isothermal phase-change with distinct phase interfaces, such regions are numerical artifacts and will be referred to as artificial mushy regions.
Alternative regularizations of phase interfaces that rely on similar parameters have been proposed in other literature \cite{danaila2014newton, zimmerman2017monolithic, rakotondrandisa2019numerical}.
These approaches often highlight that the artificial mushy region is constricted to a temperature interval around $T_m$.
Using \eqref{eq:RegularizedLiquidVolumeFraction}, that interval can be approximated as $\left[-2 \sigma, 2 \sigma\right]$, i.e. within two standard deviations of $T_m$.
It is important to note that $\sigma$ is a numerical parameter and does not carry physical meaning.
Therefore, for results in Section \ref{section:Validation}, $\sigma$ was reduced, for a given mesh, until regions where $0 < \phi_l < 1$ were small and until the shapes of the $T = T_m$ contours were not affected.

\subsection{Governing equations}

Convection-coupled phase-change can be simulated in enthalpy form \cite{voller1987enthalpy} using the following governing equations, which are balances of mass, momentum, and energy
\begin{align} 
    \label{eq:DimensionalMass}
    \nabla \cdot \mathbf{\mathpzc{u}} 
        &= 0 \\
    \label{eq:DimensionalMomentum}
    \rho_0 \left(\partial_{\mathpzc{t}} \mathbf{\mathpzc{u}} 
        + \nabla \mathbf{\mathpzc{u}} \cdot \mathbf{\mathpzc{u}}
        + \frac{\phi_s}{d} \mathbf{\mathpzc{u}}
        \right)
        + \nabla \mathpzc{p} 
        - 2 \nabla \cdot (\mu \ \mathrm{sym} \nabla \mathbf{\mathpzc{u}})
        - \rho \mathbf{g}
        &= 0 \\
    \label{eq:DimensionalEnergy}
    \partial_{\mathpzc{t}} \left(\mathpzc{C(T - T_0)}\right) 
        + \rho_l L \partial_{\mathpzc{t}} \phi_l
        + \mathbf{\mathpzc{u}} \cdot \nabla (\mathpzc{C(T - T_0)})
        - \nabla \cdot (\mathpzc{k} \nabla \mathpzc{T} )
        &= 0
\end{align}
The independent unknowns are pressure $\mathpzc{p}$, velocity $\mathbf{\mathpzc{u}}$, and temperature $\mathpzc{T}$. 
The unknown liquid volume fraction $\phi_l$, solid volume fraction $\phi_s$, and Boussinesq density $\rho$ always depend on $\mathpzc{T}$.
The thermal conductivity $\mathpzc{k}$ and volumetric heat capacity $\mathpzc{C}$ can both depend on the phase and therefore depend on $\mathpzc{T}$. 
The volume-averaged values are $\mathpzc{k} = \phi_l \mathpzc{k}_l + \phi_s \mathpzc{k}_s$ and $\mathpzc{C} = \phi_l \rho_l \mathpzc{c}_l + \phi_s \rho_s \mathpzc{c}_s$.

The mass equation \eqref{eq:DimensionalMass} is the continuity equation for an incompressible material.
The momentum equation \eqref{eq:DimensionalMomentum} is the incompressible Navier-Stokes momentum equation with Boussinesq force $\rho \mathbf{g}$ and an additional solid velocity correction \cite{wang2010comprehensive} term.
This term, $\phi_s \mathbf{\mathpzc{u}}/d$, with $d \ll 1$, relaxes the velocity toward zero in the solid region, where $\phi_s$ is the regularized solid volume fraction and $\phi_l + \phi_s = 1$.
This particular choice of solid velocity correction was used for example in \cite{brent1988enthalpy} and \cite{evans2007temporal}.
Under the Boussinesq approximation, density variations are only considered with respect to the temperature.
The incompressible flow considers a reference density $\rho_0$.
The symmetric part of the rate-of-strain tensor is denoted by $\mathrm{sym} \nabla \mathpzc{u}$.
The energy balance is in enthalpy form \cite{voller1987enthalpy}.
This invokes a reference temperature $\mathpzc{T_0}$ and produces a source term accounting for the phase-change with latent heat $L$.

\subsection{Nondimensionalization}

The independent variables were nondimensionalized with respect to characteristic scales similar to \cite{danaila2014newton}.
These include spatial scale $\mathrm{X}$, speed scale $\mathrm{U}$, and temperature scale $\delta \mathrm{T}$.
The liquid material properties were used as reference values, e.g. the liquid kinematic viscosity $\nu_l$.
Additionally, the reference temperature was chosen as the isothermal melting and freezing temperature, i.e. $\mathpzc{T}_0 = \mathpzc{T}_m$.
The nondimensional variables were chosen to be
\begin{equation}
    \label{eq:Scaling}
    \mathbf{x} = \frac{\mathbf{\mathpzc{x}}}{\mathrm{X}},
    \quad t = \frac{\mathpzc{t} \mathrm{U}}{\mathrm{X}},
    \quad p = \frac{\mathpzc{p}}{\rho_l \mathrm{U}^2},
    \quad \mathbf{u} = \frac{\mathbf{\mathpzc{u}}}{\mathrm{U}},
    \quad T = \frac{\mathpzc{T} - \mathpzc{T}_m}{\delta \mathrm{T}},
\end{equation}
In this case, the nondimensional melting and freezing temperature is always $T_m = 0$.
The phase-dependent material properties were normalized with respect to the liquid values, i.e.
\begin{align}
    \kappa &= \frac{1}{\kappa_l}\left(\kappa_s \phi_s + \kappa_l \phi_l \right) = \frac{\kappa_s}{\kappa_l} + \left(1 - \frac{\kappa_s}{\kappa_l}\right)\phi_l, 
    \label{eq:PhaseDependentConductivity}
    \\
    C &= \frac{1}{\rho_l c_l}\left(\rho_s c_s \phi_s + \rho_l c_l \phi_l \right) = \frac{\rho_s c_s}{\rho_l c_l} + \left(1 - \frac{\rho_s c_s}{\rho_l c_l}\right) \phi_l
    \label{eq:PhaseDependentVolumetricHeatCapacity}
\end{align}
In nondimensional form, the governing equations \eqref{eq:DimensionalMass}, \eqref{eq:DimensionalMomentum}, and \eqref{eq:DimensionalEnergy} then read
\begin{align} 
    \label{eq:Mass}
    \nabla\cdot\mathbf{u}
    = 0,
    \\
    \label{eq:Momentum}
    \partial_t\mathbf{u} + \nabla\mathbf{u} \cdot \mathbf{u}
    + \nabla p - \frac{2}{\mathrm{Re}} \nabla\cdot \mathrm{sym} \nabla \mathbf{u} 
    + \frac{\mathrm{Ra}}{\mathrm{Pr}} \ b \ \mathbf{\hat{g}} 
    + \frac{\phi_s}{\tau}\mathbf{u}
    = 0,
    \\
    \label{eq:Energy}
    \partial_t (CT) + \frac{1}{\mathrm{Ste}} \partial_t \phi_l + \mathbf{u} \cdot \nabla (CT) - \frac{1}{\mathrm{Re Pr}} \nabla \cdot \left( \kappa \nabla T \right) 
    = 0
\end{align}
where the Reynolds, Rayleigh, Prandtl, and Stefan numbers are defined as
\begin{equation}
    \label{eq:SimilarityParameters}
    \mathrm{Re} = \frac{\mathrm{U X}}{\nu_l},
    \quad \mathrm{Ra} = \frac{g \beta_0 \mathrm{X}^3 \delta \mathrm{T}}{\nu_l \alpha_l},
    \quad \mathrm{Pr} = \frac{\nu_l}{\alpha_l},
    \quad \mathrm{Ste} = \frac{c_l \delta \mathrm{T}}{L},
\end{equation}

In the solid velocity correction term, $\tau = d\mathrm{U}/\mathrm{X}$ is the nondimensional velocity relaxation factor.
The Boussinesq buoyancy force, with unit vector $\hat{\mathbf{g}}$ in the direction of gravity, contains a general term $b = b(T)$ which must be specified for a given liquid density model.
In cases where this density model is nonlinear, e.g. when accounting for the density anomaly observed in water, a constant reference value $\beta_0$ must be chosen for the thermal expansion coefficient.
The liquid thermal diffusivity is $\alpha_l = \kappa_l/(\rho_l c_l)$.

The nondimensional system of equations (\ref{eq:Mass}, \ref{eq:Momentum}, \ref{eq:Energy}) constitute the governing equations for the remainder of this article.
The scaling highlights the regimes of natural convection and phase-change.
Section \ref{section:Validation} presents simulations of laboratory scale experiments for the melting of octadecane and freezing of water, and describes the cases in terms of their similarity parameters.
The length and speed scales $\mathrm{X}$ and $\mathrm{U}$ are chosen separately for each simulation.
As explained in \cite{danaila2014newton}, common choices for convection-coupled melting and solidification are $\mathrm{U} = \nu_l/\mathrm{X}$ and $\mathrm{U} = \alpha_l/\mathrm{X}$.
The former yields $\mathrm{Re} = 1$ while the latter yields $\mathrm{Re} = 1/\mathrm{Pr}$.

\section{Numerical methods} \label{section:NumericalMethods}

The time-dependent problem was solved as a sequence of initial boundary values problems.
Mixed finite elements were used for the spatial discretization
and backward difference formulas were used for the temporal discretization.
Non-homogeneous Dirichlet and homogeneous Neumann boundary conditions were applied to the temperature field, while homogeneous Dirichlet boundary conditions were applied to the velocity field.
The nonlinear system was solved with Newton's method.
Reliably and accurately solving the nonlinear problem required special attention to its regularization, for which a continuation procedure was developed.

\subsection{Spatial discretization} \label{section:NumericalMethods-SpatialDiscretization}

Following a similar approach to \cite{danaila2014newton, zimmerman2017monolithic, woodfield2019stability}, mixed finite elements were used to approximate the system (\ref{eq:Mass}, \ref{eq:Momentum}, \ref{eq:Energy}) with basis functions $\psi_p, \boldsymbol{\psi}_u, \psi_T$,
yielding the weak form residual
\begin{equation} \begin{aligned}  \label{eq:WeakResidual}
    \mathcal{F} = 
    \left(\psi_p, \nabla \cdot \mathbf{u}\right)
    + \left(\boldsymbol{\psi}_u, 
        \partial_t \mathbf{u} + \nabla \mathbf{u} \cdot \mathbf{u} + \frac{\mathrm{Ra}}{\mathrm{Pr}} \ b \ \mathbf{\hat{g}} + \frac{1}{\tau}\phi_s\mathbf{u} \right)
    \\
    - \left(\nabla \cdot \boldsymbol{\psi}_u, p \right)
    + \frac{2}{\mathrm{Re}}\left(\mathrm{sym} \nabla \boldsymbol{\psi}_u, 
        \mathrm{sym}     \nabla \mathbf{u} \right)
    \\
    + \left(\psi_T, \partial_t (CT) + \frac{1}{\mathrm{Ste}} \partial_t \phi_l + \mathbf{u} \cdot \nabla (CT)  \right)
        + \frac{1}{\mathrm{Re Pr}} \left(\nabla \psi_T, \kappa \nabla T \right)
\end{aligned} \end{equation}
where $(u,v) = \int u v d\mathbf{x}$ and $(\mathbf{u},\mathbf{v}) = \int \mathbf{u}\cdot\mathbf{v} d\mathbf{x}$.

Piece-wise linear polynomials were used for approximating the pressure $p$
while piece-wise quadratic polynomials were used for the velocity $\mathbf{u}$ and temperature $T$.
This forms a mixed element which is analogous to Taylor-Hood elements for the pressure-velocity subspace.
The same element was used in \cite{woodfield2019stability},
while only $\mathbf{u}$ (and not $T$) was quadratic in \cite{danaila2014newton} and \cite{zimmerman2017monolithic}.

Homogeneous Dirichlet boundary conditions were always applied to the velocity, which reflect a stationary solid whenever it connects to the boundary, and a no-slip boundary for the liquid phase.
Both non-homogeneous Dirichlet and homogeneous Neumann boundary conditions were applied to the temperature.
No boundary conditions were applied to the pressure.
Therefore, the pressure solution was only defined up to a constant.
Successful solution required informing the linear solver of the appropriate nullspace, which was automated by the software in Section \ref{section:Firedrake}.
To solution was made unique by subtracting the mean pressure (i.e. $\int p \ d\mathbf{x}$), which also therefore guaranteed zero mean pressure in the post-processed solution.

\subsection{Temporal discretization}
\label{section:NumericalMethods-TemporalDiscretization}

The time derivatives were discretized with the BDF2 method, i.e. the second order method from the family of $k^{\mathrm{th}}$ order constant time step size ($\Delta t$) backward difference formulas (BDF-$k$) \cite{ascher1998computer}.
For some unknown $w^n$ at time step $n$, the BDF-$k$ discretization can be written as
\begin{equation} \label{eq:BDF}
    \partial_t w = \frac{1}{\Delta t}\sum_{i = 0}^{k} a_i w^{n - i}
\end{equation}
where the coefficients $a_i$, shown in Table \ref{table:BDFCoefficients} depend only on $k$.
This requires storing solutions from the previous $k$ time steps. 
\begin{table}
    \centering
    \begin{tabular}{c|cccccc}
        $k$ & $a_0$ & $a_1$ & $a_2$ & $a_3$ \\
        \hline
        1 & 1 & -1 & & & \\
        2 & 3/2 & -2 & 1/2 & & \\
        3 & 11/6 & -3 & 3/2 & -1/3
    \end{tabular}
    \caption[.]{Coefficients for backward difference formulas up to third order, taken from \cite{ascher1998computer}.
        For example, the backward Euler method is given by $k = 1$, and BDF2 is given by $k = 2$.}
    \label{table:BDFCoefficients}
\end{table}
In addition to the independent unknowns, (\ref{eq:BDF}) was applied directly to $\phi_l$ and to $CT$ in (\ref{eq:Energy}).
For example, applying the second order formula (BDF2) to $\phi_l$ yields $\partial_t \phi_l = (3\phi_l(T^n) - 4\phi_l(T^{n - 1}) + \phi_l(T^{n - 2}))/\left(2 \Delta t\right)$.

\subsection{Nonlinear solution}
\label{section:NonlinearSolution}

Many terms in the residual (\ref{eq:WeakResidual}) are highly nonlinear.
The nonlinear system was solved with Newton's method as follows.
Consider the vector-valued system solution $\mathbf{w}^n = \begin{pmatrix} p^n & \mathbf{u}^n & T^n \end{pmatrix}^\mathsf{T}$ for any discrete time step $n$.
The nonlinear problem's solution is approximated for the next time step by solving a sequence of linear problems
\begin{equation} \label{eq:NewtonMethod}
    D_\mathbf{w} \mathcal{F}(\mathbf{w}_j)(\mathbf{w}_{j + 1} - \mathbf{w}_j) 
        = \mathcal{F}(\mathbf{w}_j),
\end{equation}
whose iterates $\mathbf{w}_{j + 1}$ converge to $\mathbf{w}^n$.
Here, $D_\mathbf{w} \mathcal{F}$ stands for the Gateaux derivative of $\mathcal{F}$.
It can be derived analytically, which was done for a similar problem in \cite{zimmerman2017monolithic}.
Or as done in the current work, it can be symbolically computed using the software in Section \ref{section:ImplementationAndVerification}.
Each iteration of \eqref{eq:NewtonMethod} is a linear system which can be solved robustly with a direct solver.
At each new time step $n$ with unknown solution $\mathbf{w}^n$, the initial guess for the nonlinear solver was first taken as the known solution $\mathbf{w}^{n - 1}$ from the previous time step $n - 1$.

It is not simple to generate suitable initial guesses to reliably converge the nonlinear solver.
Convergence is particularly difficult when using a small enough regularization parameter $\sigma$ in (\ref{eq:RegularizedLiquidVolumeFraction}) such that the solution is not highly sensitive to the regularization.
A continuation procedure, detailed in the next section, was developed to automatically search for better initial guesses when the nonlinear solver fails. 
The key idea of the procedure is to recognize that the regularity of nonlinear problem's solution space is dominated by $\sigma$.
Only $\sigma$, no other aspects of the problem, was modified during continuation.
No data from previously solved time steps was modified.
Entirely automated as part of the implementation in Section \ref{section:ImplementationAndVerification}, the procedure successfully solved multiple benchmark problems on coarse uniform meshes with large time step sizes using small $\sigma$.

\subsection{Continuation}
\label{section:Continuation}

\begin{figure}[h]
    \centering
    \includegraphics[width=0.8\linewidth]{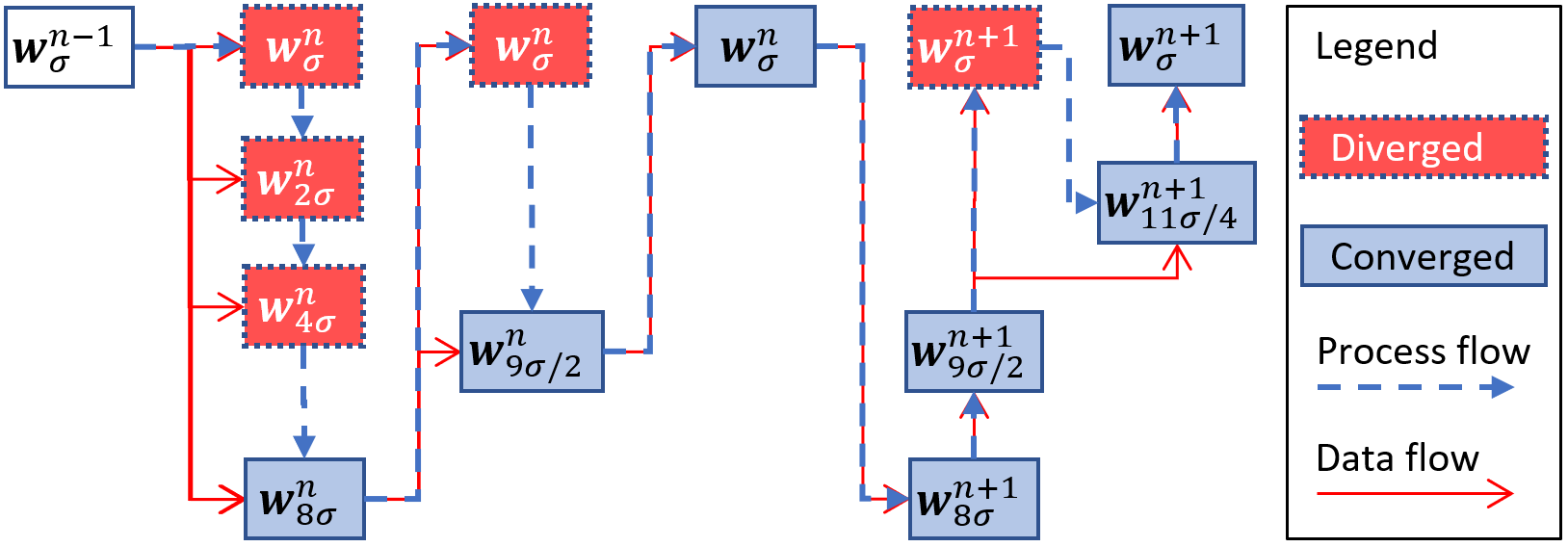}
    \caption[.]{
        Example process flow for the continuation procedure.
        
        Newton's method was used to solve a nonlinear problem $\mathcal{F}_\sigma(\mathbf{w}_\sigma^n) = 0$ for $\mathbf{w}^n_\sigma$ with regularization parameter $\sigma$ at time $n$.
        First, the known solution from the previous time step, $\mathbf{w}^{n-1}_\sigma$, was used as an initial guess.
        The solver diverged.
        The solution was reinitialized and $\sigma$ was recursively doubled (i.e. the unbounded search mode), until the solver converged when using $8\sigma$.
        This provided an intermediate solution, $\mathbf{w}^n_{8\sigma}$, which replaced the initial guess.
        The solver still failed with the new guess.
        Therefore, the next attempted $\sigma$ bisected the target value $\sigma$ and the nearest successful value $8 \sigma$ (i.e. the bounded search mode).
        For the initial guess $\mathbf{w}^n_{9\sigma/2}$, the solver converged.
        The simulation then proceeded to the next time step.
        It was attempted to use the sequence of $\sigma$ values from the previous time step was re-used in order to avoid costly solver failures.
        The bounded search mode had to be applied once more, adding an additional intermediate solution to the sequence.
        }
    \label{fig:continuation}
\end{figure}

Initial guesses were improved by solving intermediate problems with an increased value of $\sigma$.
Increasing $\sigma$ improves the regularity of the solution, making the solver more likely to converge even for a poor-quality initial guess.
Whenever a solution was found, it was used as an initial guess for solving the same system again, albeit with a smaller value of $\sigma$.
The challenge hence was to find a sequence of $\sigma$ values resulting in solvable problems, with the final solution using the originally specified $\sigma$.
The search process had two modes:
Either the search was bounded by two values of $\sigma$ or it was unbounded.
In the unbounded case, $\sigma$ was always doubled until the problem could be solved.
In the bounded case, new $\sigma$ values were always inserted into the sequence to bisect the nearest solved and unsolved values.
Figure \ref{fig:continuation} shows a detailed example of this process.

When applying this approach, the resulting sequence of intermediate $\sigma$ values depends on the exact nonlinear solver method used, including on its options such as convergence criteria and maximum iterations.
It was found that using the relative residual as a convergence criterion would allow for significant accumulation of error in the absolute residual between time steps.
For this reason, the relative tolerance convergence criterion was disabled, forcing the nonlinear solver to reach a low absolute residual or otherwise fail.
Any single call to the nonlinear solver was only allowed twenty-four maximum iterations to reach the convergence criteria, meaning again that the solver would otherwise report failure.

\section{Implementation and verification with open-source libraries} \label{section:ImplementationAndVerification}

In order to reduce the required coding effort, leverage state-of-the-art methods and solvers, and increase the usability of the software and reproducibility of results, the code was implemented using the open source finite element library Firedrake \cite{rathgeber2016firedrake}.
On top of Firedrake, for this work, the open source Sapphire \cite{zimmerman2020sapphire} framework was developed which implements the methods from Section \ref{section:NumericalMethods}.
Sapphire also provides an automated procedure for verifying implementations with the method of manufactured solutions (MMS).
In addition to a general overview of Firedrake and Sapphire, the results of applying Sapphire's MMS procedure to verify the current implementation are also presented here.

\subsection{Firedrake} \label{section:Firedrake}

Firedrake \cite{rathgeber2016firedrake} is a Python package which automates many aspects of finite element method software development.
It uses the Unified Form Language (UFL) \cite{alnaes2012unified} to provide users with an abstract interface for rigorously specifying variational problems and their finite element discretizations.
With the problem and discretization specified by the user, Firedrake employs the Two-Stage Form Compiler (TSFC) \cite{homolya2017tsfc} to automatically write efficient code for the finite element assembly.
The linear and nonlinear solvers are provided by PETSc \cite{petsc-user-ref, petsc-efficient} via PETSc for Python \cite{Dalcin2011}.
For the results in this work, all linear systems were solved with MUMPS \cite{MUMPS01, MUMPS02, PTSCOTCH}, and all nonlinear systems were solved with PETSc's SNES (Scalable Nonlinear Equations Solvers) line search method, both via the Firedrake interfaces.

By default, Firedrake automatically determines the quadrature degree used during finite element assembly for a given problem.
Alternatively, the user can specify a degree.
In the current work, when integrating the regularized liquid volume fraction \eqref{eq:RegularizedLiquidVolumeFraction} with small regularization parameter value $\sigma$, Firedrake  automatically determined excessively large quadrature degrees.
Computations were performed more efficiently by reducing the quadrature degree, which reduced the number of function evaluations needed for each degree of freedom when performing finite element assembly.
This essentially introduced a new numerical parameter to the model, the quadrature degree, $q$.
Therefore, $q$ is included in the sensitivity studies of Section \ref{section:Validation}.

Firedrake depends on a stack of open source scientific software libraries.
A benefit is that ongoing technological developments from a variety of open source scientific software projects are leveraged.
A downside is that the software can be difficult to configure and compile.
To manage this complexity, the Firedrake project maintains an installation script.
Furthermore, to facilitate the reproduction of scientific results, Firedrake provides a command line tool which installs a specified set of versions and configurations throughout the software stack.
The exact versions of Firedrake and its software stack used for the results in the current work were archived with a DOI on Zenodo \cite{zenodo/Firedrake-20200611.3}, ensuring the reproducibility of these results.

\subsection{Sapphire} \label{section:Sapphire}

Sapphire is a Python package that was developed for the purposes of the current work.
Sapphire provides a Python class for time dependent simulations that are governed by partial differential equations (PDEs).
The PDEs must be discretized in space with finite elements and in time with finite differences.
Spatial discretization is handled by Firedrake, with the PDE being defined in variational form using UFL.
The time discretization module includes backward different formulas, as written in Section \ref{section:NumericalMethods-TemporalDiscretization}, up to sixth order.
Five arguments are required to instantiate a Sapphire simulation: 
\begin{itemize}
    \itemsep0em
    \item a mesh, e.g. from Firedrake's built-in mesh functions or converted from an external meshing library, 
    \item a finite element or mixed finite element defined by UFL, 
    \item the residual of the governing equations in variational form, e.g. (\ref{eq:WeakResidual}), defined by UFL, 
    \item a (possibly empty) list of Firedrake Dirichlet boundary conditions, 
    \item initial values as a Firedrake finite element function.
\end{itemize}
With these defined, a simulation is ready to run.

The solver's accuracy can be verified using the method of manufactured solutions (MMS), e.g. as described in \cite{roache2002code}.
Manually deriving MMS source terms creates many opportunities for errors, especially for systems of PDEs involving vector calculus.
Sapphire provides a MMS module which uses UFL to automatically derive the source terms and augment the variational problem which needs to be solved for verification.
As part of its test suite, Sapphire uses the MMS module to verify the accuracy of its solutions to many different PDEs.

Sapphire is a flexible tool, is publicly available, and is freely licensed.
Simulations have been implemented in Sapphire which solve the 
heat equation, 
convection-diffusion, 
steady and unsteady incompressible Navier-Stokes, 
natural convection, 
phase-change in enthalpy form without convection,
and convection-coupled phase-change in enthalpy form,
the last of which is the focus of this article.
The version of Sapphire used to produce the results in the current work was archived with a DOI on Zenodo at \cite{zimmerman2020sapphire}.

\subsection{Solver verification} \label{section:MMS}

Numerical solutions to the weak residual (\ref{eq:WeakResidual}) approximate solutions to the governing equations (\ref{eq:Mass}, \ref{eq:Momentum}, \ref{eq:Energy}).
As there is no sufficiently complex analytical reference solution available, spatial and temporal convergence were verified via the method of manufactured solutions (MMS).

\subsubsection{MMS procedure}
MMS is a general method for verifying PDE solvers.
Manufactured solutions do not necessarily need to have physical interpretations \cite{roache2002code}.
Their purpose purpose rather is to reveal coding errors and to verify the solver.
Therefore, a manufactured solution can be any function which is sufficiently differentiable in space and time to exercise the spatial and temporal derivatives in the governing equations.
There is an additional benefit to MMS when approximating solutions to strong form governing equations via weak formulations.
In addition to verifying the code, the weak formulation is also verified, because the weak form is tested against the strong form.

Substituting a manufactured solution, which is typically not an exact solution, into the governing equations yields a source term in each equation.
Adding the source terms to the governing equations creates an augmented problem to which the manufactured solution is an exact solution.
To verify a solver, it is applied to the augmented problem, and the resulting solution must approximate the manufactured solution with the expected orders of accuracy.

To apply the MMS procedure for verifying a solver's spatial or temporal orders of accuracy, Sapphire requires the user to
\begin{itemize}
    \itemsep0em
    \item implement the strong form governing equations (\ref{eq:Mass}, \ref{eq:Momentum}, \ref{eq:Energy}) with UFL,
    \item implement a manufactured solution with UFL,
    \item and set a single time step size $\Delta t$ and a list of meshes with decreasing cell sizes $h$ for spatial verification, or a single mesh and a list of $\Delta t$ for temporal verification.
\end{itemize}
Separate manufactured solutions can be used to independently verify the spatial and temporal discretizations.

The strong form governing equations and manufactured solutions are defined in a Python module, and the verification process is run with a Python script.
Then, the module is passed to one of Sapphire's MMS verification functions.
Sapphire automatically 
\begin{itemize}
    \itemsep0em
    \item derives the source term for each component of the manufactured solution, multiplies them with  appropriate test functions, and adds them to the weak form,
    \item sets initial values and boundary conditions as given by the solution,
    \item runs simulations for specified lists of $h$ or $\Delta t$,
    \item tabulates errors between the approximate finite element solutions and the manufactured solutions,
    \item reports convergence rates with respect to $h$ or $\Delta t$,
    \item and asserts that the reported convergence rate matches the expected order of accuracy within a specified tolerance.
\end{itemize}

\subsubsection{Verification of convection-coupled phase-change}
Verification of the current convection-coupled phase-change solver was performed on a unit square domain.
Dirichlet boundary conditions were applied to the velocity and temperature.
The manufactured solution was
\begin{align*}
    \mathbf{u}_M &= \exp{(t/2)}\sin{(2\pi x)}\sin{(\pi y)} \hat{\mathbf{i}} + 
        \exp{(t/2)} \sin{(\pi x)} \sin{(2 \pi y)}\hat{\mathbf{j}}, \\
    p_M &= p^* - \int_\Omega p^* d\mathbf{x}, \quad p^* = -\sin{(\pi x - \pi/2)} \sin{(2 \pi y - \pi/2)} \\
    T_M &= 0.5 \sin{(2 \pi x)} \sin{(\pi y)} \left(1 - \exp{(-t^2/2)}\right)
\end{align*}
where $x$ and $y$ are the components of the position $\mathbf{x}$.
The definitions of $p^*$ and $p_M$ were separated so that $p^*$ could be chosen freely while $p_M$ would maintain zero mean.
For the buoyancy term in (\ref{eq:Momentum}), a classical linear Boussinesq model was used by setting $b = T$.
The similarity parameters were set to Re = 20, $\mathrm{Ra} = 2.5 \times 10^6$, $\mathrm{Pr} = 7.0$, and $\mathrm{Ste} = 0.13$. 
The material property parameters were set to $\rho_s/\rho_l = 0.92$, $c_s/c_l = 0.50$, and $\kappa_s/\kappa_l = 3.8$.
The numerical parameters were set to $\sigma = 0.1$, $\tau = 10^{-6}$, and $q = 4$.

To verify spatial discretization accuracy independently of any time discretization error, the manufactured solution was evaluated at time $t = 1$, and no further time dependency was assumed.
The resulting problem was solved in a single pseudo-time step.
This was repeated on a series of uniformly refined meshes.
To independently verify the temporal discretization accuracy, a sufficiently refined mesh was used such that the spatial discretization error would not dominate the total error.
The unsteady problem was solved repeatedly with a sequence of time step sizes $\Delta t$.
Table \ref{table:MMSVerification} shows the observed convergence orders with respect to $h$ and $\Delta t$.
As expected, the temperature and velocity solutions converge quadratically.
On the other hand, the pressure error shows super-convergence.
Given that the pressure solution is not of primary interest in this work, its super-convergence was not further investigated.

\begin{table}
    \centering
    \begin{tabular}{c|cccccc}
        $h$ 
        & $\lVert p_h - p_M \rVert _{\mathrm{L^2}}$ & $r_{h,p}$ 
        & $\lVert \mathbf{u}_h - \mathbf{u}_M \rVert _{\mathrm{H^1}}$ & $r_{h,\mathbf{u}}$ 
        & $\lVert T_h - T_M \rVert _{\mathrm{H^1}}$ & $r_{h,T}$ 
        \\
        \hline
        1/32  &	4.748e-01 &	      &	1.910e-02 &	      &	1.534e-03 &	     \\
        1/64  &	3.041e-02 &	3.965 &	4.537e-03 &	2.074 &	3.791e-04 &	2.017 \\
        1/128 &	1.912e-03 &	3.991 &	1.120e-03 &	2.018 &	9.447e-05 &	2.005 \\
        1/256 & 1.205e-04 & 3.988 &	2.796e-04 &	2.002 &	2.360e-05 &	2.001
    \end{tabular}
    \begin{tabular}{c|cccc}
        $\Delta t$ 
        & $\lVert \mathbf{u}_h - \mathbf{u}_M \rVert _{\mathrm{H^1}}$ & $r_{\Delta t,\mathbf{u}}$ 
        & $\lVert T_h - T_M \rVert _{\mathrm{H^1}}$ & $r_{\Delta t,T}$ 
        \\
        \hline
        1/4   & 3.468e-02	&       & 9.923e-03	&       \\
        1/8   & 1.087e-02	& 1.674	& 3.330e-03	& 1.575 \\
        1/16  & 2.858e-03	& 1.927	& 8.876e-04	& 1.908 \\
        1/32  & 7.167e-04	& 1.996	& 2.202e-04	& 2.011
    \end{tabular}
    \caption[.]{
        Empirical convergence of second-order spatial and temporal discretizations.
        
        The errors for discrete pressure $p_h$, velocity $\mathbf{u}_h$, and temperature $T_h$ are computed in their natural norms with respect to the manufactured solutions $p_M$, $\mathbf{u}_M$, and $T_M$ to compute convergence rates $r_{h,p}$, $r_{h,\mathbf{u}}$, and $r_{h,T}$.
        The time discretization errors are computed in the $L^2$-norm to determine convergence rates $r_{\Delta t,\mathbf{u}}$ and $r_{\Delta t,T}$.
    }
    \label{table:MMSVerification}
\end{table}

\section{Validation with benchmark experiments}
\label{section:Validation}

\begin{figure}[h]
    \centering    
    \includegraphics[width=0.55\linewidth]{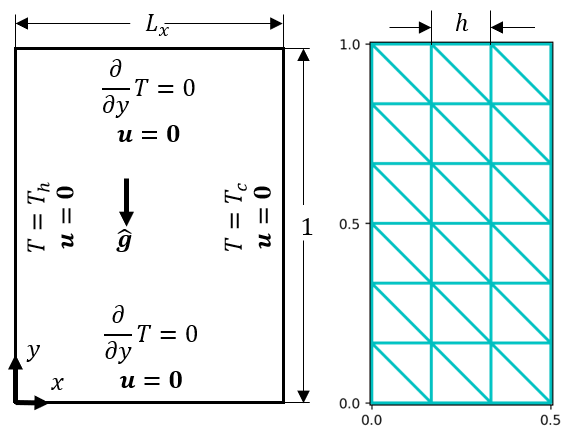}
    \caption[.]{
        The geometric domain, boundary conditions, and mesh.
        
        Left)
        All computations used a rectangular domain with unit height and width $L_x$.
        Most used the unit square (i.e. $L_x = 1$).
        The left wall was kept constant at a higher temperature, $T_h$.
        The right wall was kept constant at a lower temperature, $T_c$.
        The top and bottom walls were adiabatic.
        All walls had no slip conditions on the velocity.
        The direction of gravity for the momentum equation \eqref{eq:Momentum} was $\mathbf{\hat{g}} = (0, -1)^\mathrm{T}$.
        Right) An example uniform triangular mesh of a rectangular domain with $L_x = 1/2$.
        The cell edge length is $h = 1/6$.
        }
    \label{fig:TestDomainAndMesh}
\end{figure}

Two experimental data sets from the literature were considered in order to validate the implemented model.
Simulations were run for each, with the goal of accurately modeling the time evolution of the phase interface, i.e. the $T = T_m$ isotherm.
First, octadecane melting was simulated for comparison to an experiment from \cite{okada1984analysis}.
Second, water freezing was simulated for comparison to experiments from \cite{kowalewski1999freezing}.
Both cases used a unit square geometric domain with boundary conditions shown in Figure \ref{fig:TestDomainAndMesh}.
The top and bottom boundaries were assumed to be adiabatic, while the left and right walls were kept respectively at constant hot temperature $T_h$ and cold temperature $T_c$.
Zero velocity boundary conditions were applied on each wall.
For both benchmarks, gravity points vertically downward, i.e. $\mathbf{\hat{g}} = (0, -1)^\mathrm{T}$.
Also in both cases, a uniform triangular mesh with cell edge length $h$ was used.
An example mesh with $h = 1/6$ is shown in Figure \ref{fig:TestDomainAndMesh}, on the right.

Each of the two following subsections is devoted to one of the two cases.
In both cases, the subsection includes
\begin{enumerate}
    \itemsep0em
    \item a description of the experimental setup.
    \item a description of the simulation setup. This includes the scaling and similarity parameters, the phase-dependent material properties, the buoyancy model, the procedure for obtaining initial values, and the boundary condition values.
    \item a sensitivity study involving the mesh cell size $h$, time step size $\Delta t$, phase interface regularization parameter $\sigma$, solid velocity relaxation parameter $\tau$, and quadrature degree $q$.
    With respect to a nominal set of parameters values, each parameter is varied individually and the resulting phase interfaces are compared at a fixed time.
    \item selected simulation results compared to experimental data and to corresponding simulation results in \cite{rakotondrandisa2020toolbox}.
\end{enumerate}
Where compute times are reported, the simulations ran on a single core of an Intel(R) Core(TM) i5-7500 CPU.
The exact versions of Sapphire and Firedrake used to generate all results in this section are archived respectively at \cite{zimmerman2020sapphire} and \cite{zenodo/Firedrake-20200611.3}.

\subsection{Melting octadecane}
\label{subsection:Validation_Melting_octadecane}

The melting of octadecane was simulated for comparison to an experiment from \cite{okada1984analysis}.
In the experiment, a slab of octadecane was insulated on the top and bottom.
The material was initially at its melting temperature.
One side was heated to melt the material.

\subsubsection{Simulation set-up}

The speed scale was chosen as $\mathrm{U} = \nu_l/X$.
The height of the experimental cavity as taken from \cite{okada1984analysis} was used as the length scale, i.e. $\mathrm{X} = 0.015 \ \mathrm{m}$.
A realistic value for the liquid kinematic viscosity, $\nu_l = 10^{-5} \ \mathrm{m^2/s}$ as taken from \cite{wang2010comprehensive}, furthermore yields the characteristic time scale $\mathrm{X}^2/\nu_l = 22.5 \ \mathrm{s}$.
Based on physical parameters provided in \cite{danaila2014newton} and according to the previously introduced definitions \eqref{eq:SimilarityParameters}, the similarity parameters are Ra~=~3.27$\times10^5$, Pr~=~56.2, and Ste~=~0.0450. 
Octadecane has approximately constant volumetric heat capacity and thermal conductivity between phases.
Therefore the normalized values for the energy balance \eqref{eq:Energy} are $C = \kappa = 1$.
Finally, a linear Boussinesq model with constant thermal expansion coefficient is assumed.
In this case, the buoyancy function in the nondimensional momentum balance \eqref{eq:Momentum} is simply $b = T$. 

For the nondimensional temperature corresponding to the melting point, i.e. $T = T_m = 0$, the regularized liquid volume fraction as defined in (\ref{eq:RegularizedLiquidVolumeFraction}) is evaluated to be $\phi_l = 0.5$.
This is true for any chosen regularization parameter $\sigma$.
In order to capture the experimental conditions, in which the material at time zero had been solid with a temperature close to the melting temperature, the initial temperature for the simulation was offset by a small amount, i.e. $T_0 = -0.01$.
This approach follows earlier work, e.g. \cite{danaila2014newton, zimmerman2017monolithic, rakotondrandisa2019numerical}.
For this particular choice of the initial temperature $T_0$, setting $\sigma = 0.004$ yields an initial liquid volume fraction $\phi_l = 0.006$, meaning that more than 99\% of the material is initially solid. 

During the experiment, the left boundary was kept at a constant temperature well above the melting point, while the right boundary was kept at a temperature close to the melting point. 
For the simulation boundary conditions shown in Figure \ref{fig:TestDomainAndMesh}, this translates into $T_h = 1$ and $T_c = T_0$.
The high temperature at the left boundary causes melting from left to right.

\subsubsection{Sensitivity to numerical parameters}

\begin{figure}[h]
    \centering
    \includegraphics[width=1.\linewidth]{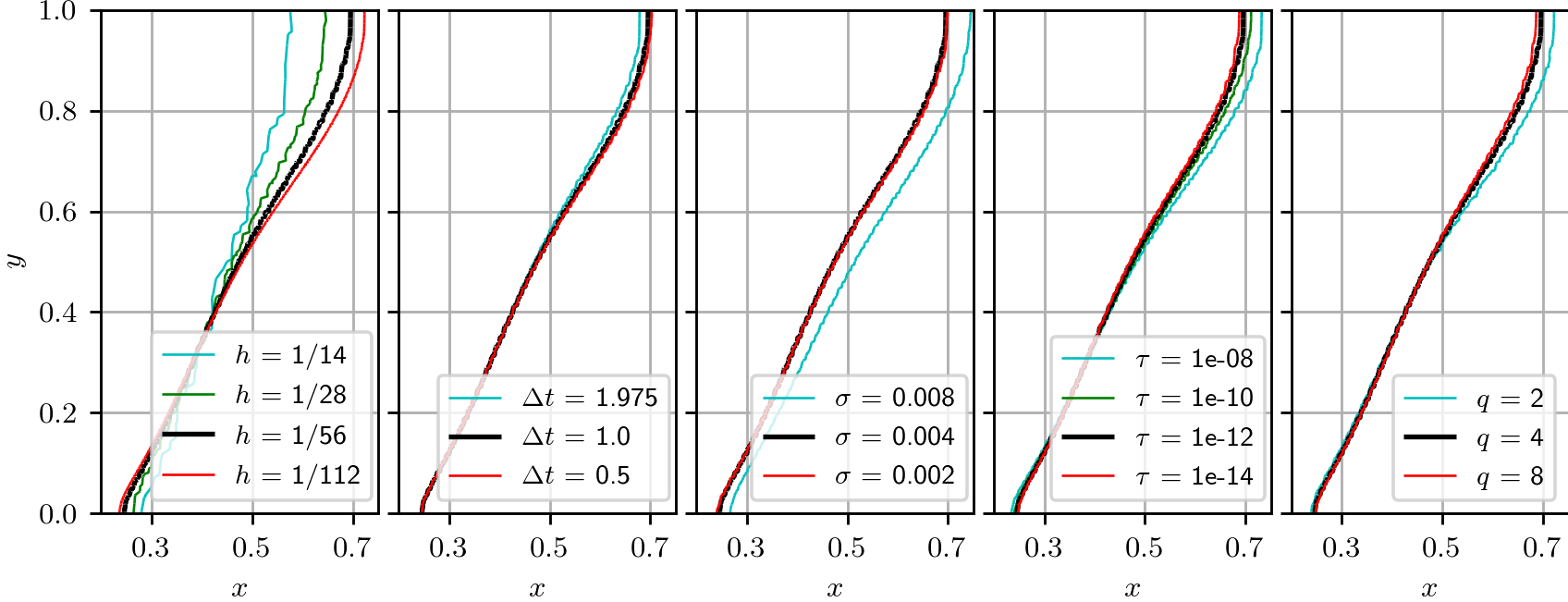}
    \caption[.]{
        Sensitivity study for the octadecane melting simulation.
        
        The numerical parameters in this study were the mesh cell size $h$, time step size $\Delta t$, phase interface regularization parameter $\sigma$, solid velocity relaxation parameter $\tau$, and quadrature degree $q$.
        With respect to nominal values ($h = 1/56$, $\Delta t = 1$, $\sigma = 0.004$, $\tau = 10^{-12}$, and $q = 4$, each parameter was varied individually and the resulting phase interfaces were compared at time $t = 79$.
        In each plot, the nominal result is shown in black.
        }
    \label{fig:Sensitivity_OctadecaneMelting}
\end{figure}

For this sensitivity study, the nominal parameter values were $h = 1/56$, $\Delta t = 1$, $\sigma = 0.004$, $\tau = 10^{-12}$, and $q = 4$.
From their nominal values, each parameter was varied independently.
The resulting phase interfaces were compared at time $t = 79$.
The results are shown in Figure \ref{fig:Sensitivity_OctadecaneMelting}.
The time $t = 79$ is also the final simulated time presented in Figure \ref{fig:Validation_OctadecaneMelting}. 

The mesh cell size, $h$, was varied from a maximum of $h = 1/14$ to a minimum of $h = 1/112$.
The upper part of the phase interface was most sensitive to $h$.
Between the smallest values, $h = 1/56$ and $h = 1/112$, the position of the phase interface in the upper portion of the domain still visibly changed; but overall there appeared to be an asymptotic convergence behavior.
The time step size, $\Delta t$, was varied from a maximum of $\Delta t=1.975$ to a minimum of $\Delta t=0.5$.
For the largest size, $\Delta t = 1.975$, there was a significant decrease in the melting near the top wall.
For $\Delta t \le 1$, the result was no longer significantly changed by reducing $\Delta t$.
The regularization parameter, $\sigma$, was varied from a maximum of $\sigma = 0.008$ to a minimum of $\sigma = 0.002$.
For the largest value, $\sigma = 0.008$, the entire phase interface shifted rightward.
This was because, at the initial temperature $T_0 = -0.01$, the initial liquid volume fraction was $\phi_l = 0.1056$.
This meant that about ten percent of the material was already melted at the initial time,
shifting the entire phase interface in the direction of melting.
For $\sigma = 0.004$, less than one percent of the material was melted at the initial time, and this effect no longer appeared to dominate.
Between the two smallest values, $\sigma = 0.004$ and $\sigma = 0.002$, the change is barely visible.
The solid velocity relaxation parameter, $\tau$, was varied from a maximum of $\tau = 10^{-8}$ to a minimum of $\tau = 10^{-14}$.
For the largest value, $\tau = 10^{-8}$, significantly more melting occurred near the top wall.
An asymptotic behavior was observed while reducing $\tau$ by orders of magnitude.
The change when reducing from $\tau = 10^{-12}$ to $\tau = 10^{-14}$ was negligible.
The quadrature degree, $q$, was varied from a minimum of $2$ to a maximum of $8$.
For the smallest value, $q = 2$, significantly more melting occurred near the top wall.
Between $q = 4$ and $q = 8$ the change was negligible.

\subsubsection{Results and discussion}

\begin{figure}[h]
    \includegraphics[width=1.\linewidth]{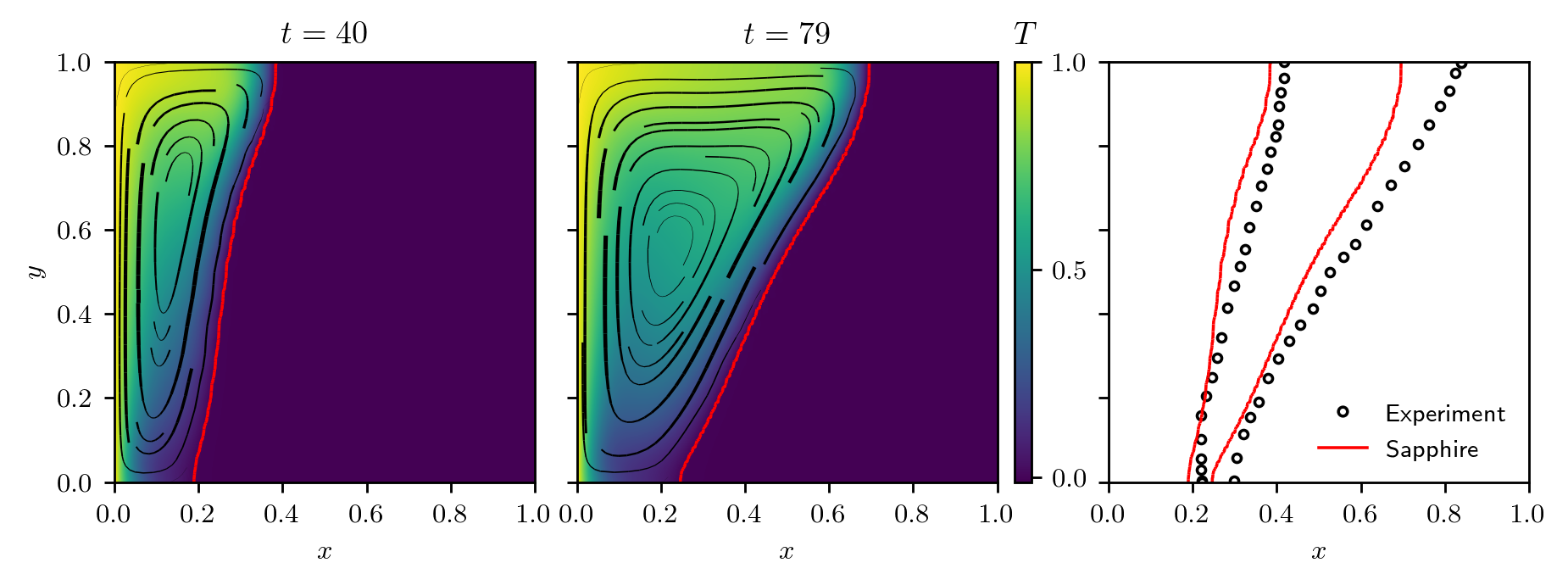}
    \caption[.]{
        Octadecane melting simulation results compared to an experiment from \cite{okada1984analysis}.
        
        A slab of octadecane, initially at its melting temperature, was insulated on the top and bottom and higher temperature was applied on the left wall to initiate melting.
        Left and Center) Simulation results with colored temperature field and black velocity  streamlines (having thickness proportional to their local speed), and the liquidus isotherm indicating the phase interface in red.
        Right) A comparison of the phase interfaces between the experiment and simulation at both times.
        }
    \label{fig:Validation_OctadecaneMelting}
\end{figure}

Figure \ref{fig:Validation_OctadecaneMelting} shows the temperature field, streamlines of the velocity field, and the post-processed position of the phase interface at times $t = 40$ and $t = 79$.
The numerical parameters were $h = 1/56$, $\Delta t = 1$, $\sigma = 0.004$, $\tau = 10^{-12}$, and $q = 4$.
Convection in the liquid phase caused a sharper temperature gradient near the top of the phase interface than towards its bottom.
This in turn caused a faster melting rate in the upper region.
This effect was more pronounced at the later time.

At the final time ($t = 79$), after a total of seventy-nine time steps ($\Delta t~=~1$), 2847 total Newton iterations were used which ran altogether in a total of one hour (wall-time).
This averages to about thirty-six Newton iterations per time step.
This includes all iterations from intermediate problems during the continuation procedure sketched in Figure \ref{fig:continuation}.
For all time steps, the largest intermediate regularization parameter value was $\sigma = 0.256$. 
For the first time step, only a single intermediate value, $\sigma = 0.016$, was required.
By the final time, the intermediate values were ($\sigma$ = 0.256, 0.13, 0.067, 0.0355).

The simulation under-predicts melting with respect to the experiment.
Qualitatively, the largest deviation is near the top wall at the end of the simulation.
The previously conducted sensitivity study, shown in Figure \ref{fig:Sensitivity_OctadecaneMelting}, suggests that the phase interface near the top wall could agree better with the experimental data by reducing the mesh cell size $h$, but no such improvement should be expected near the bottom wall or interior.

It has already been reported in \cite{rakotondrandisa2019numerical} that there is some uncertainty about the insulation of the top and bottom walls in the experiment.
The phase interface as observed in the experiment is clearly not orthogonal to the top wall.
A further analysis of the impact of various formulations for the boundary conditions is outside of the scope of the current work.
Qualitatively, especially away from the top wall, the results compare well to the experiment.
One possible way forward could be to apply a heat flux boundary condition on the top wall instead of the adiabatic boundary condition which was used.
For further validation, the experimental conditions should be examined in more detail, especially with regards to the boundary conditions.

\begin{figure}[h]
    \includegraphics[width=1.\linewidth]{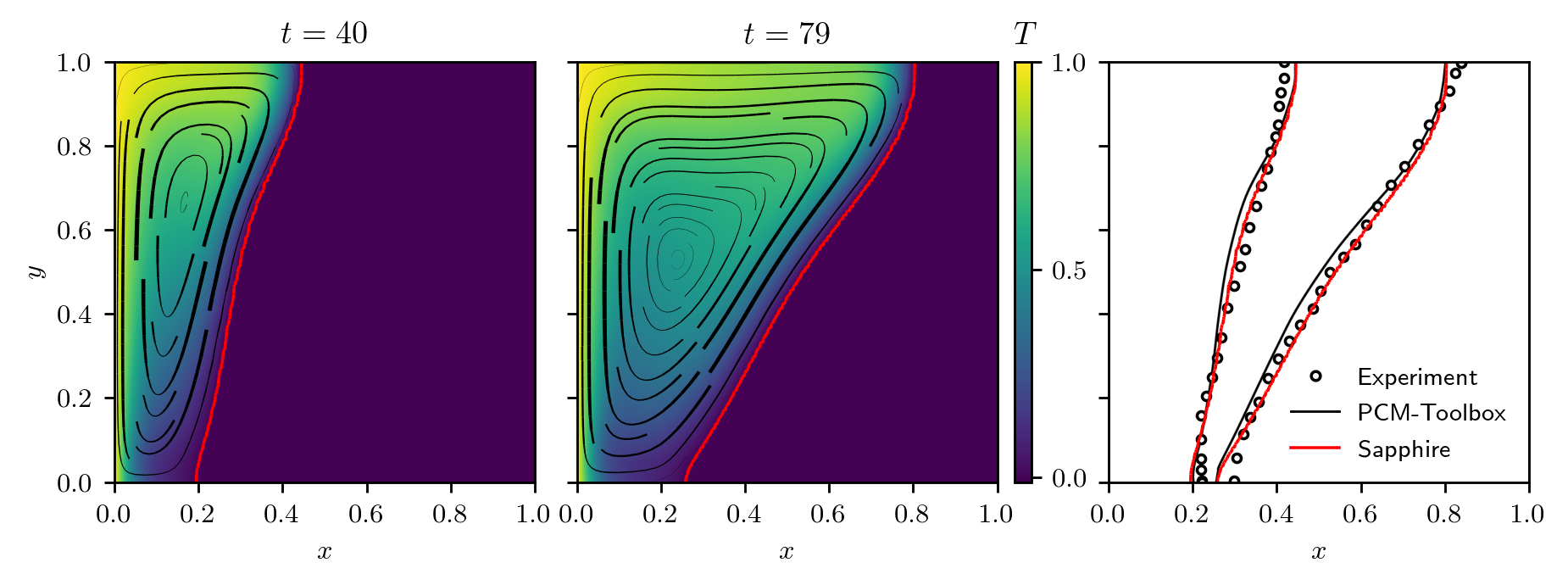}
    \caption[.]{
        Comparison to octadecane melting simulation in \cite{rakotondrandisa2020toolbox} and experiment from \cite{okada1984analysis}.
        
        Compared to the result in Figure \ref{fig:Validation_OctadecaneMelting}, only $\sigma$ was adjusted (to $\sigma = 0.00875$) in order to approximate the regularization from \cite{rakotondrandisa2020toolbox}.
        }
    \label{fig:Validation_OctadecaneMelting_CompareToPCMToolbox}
\end{figure}

It is also possible to better match the experiment by adjusting the regularization parameter $\sigma$ and velocity relaxation factor $\tau$.
This also allows a more direct comparison between the current approach and the latest results from \cite{rakotondrandisa2020toolbox}.
Therefore, another Sapphire simulation was run for the current case, only changing $\sigma$ and $\tau$ from the result in Figure \ref{fig:Validation_OctadecaneMelting}.
Figure \ref{fig:Validation_OctadecaneMelting_CompareToPCMToolbox} compares the result to \cite{rakotondrandisa2020toolbox} and to the experiment.
The simulation from \cite{rakotondrandisa2020toolbox} used 2900 mesh cells which were adaptively re-meshed at every time step, clustering cells near the phase interface.
Sapphire's simulation used 3136 uniformly distributed mesh cells (i.e. $h = 1/56$).
To approximate the regularization in \cite{rakotondrandisa2020toolbox}, $\sigma$ was set to $\sigma = 0.00875$.
Similar to the approach \cite{rakotondrandisa2020toolbox}, simulations were run with $10^{-6} \le \tau \le 10^{-8}$.

Figure \ref{fig:Validation_OctadecaneMelting_CompareToPCMToolbox} shows the result with $\tau = 10^{-8}$.
The result compares very well to \cite{rakotondrandisa2020toolbox} and to the experiment.
A total of 949 Newton iterations were used and the simulation required twenty-five minutes of compute time.
A run time of one hour and nine minutes was reported in \cite{rakotondrandisa2020toolbox}.
Their smaller time step size of $\Delta t = 0.1$ could account for their longer run time.
Many more computer hardware and software variables would need to be controlled to make a detailed performance comparison.
Still, it is remarkable that similar results were found with uniform and adapted meshes using similar number of mesh cells and the same finite elements.

\subsection{Freezing water}
The freezing of distilled water was simulated for comparison to benchmark experiments in \cite{kowalewski1999freezing}, where a cube of liquid water was frozen from one side.
For water-ice, the density, heat capacity, and thermal conductivity vary significantly between phases.
Furthermore, liquid water's thermal expansion coefficient is nonlinear, with the sign inverting at the temperature of water's greatest density.
This is sometimes referred to as the density anomaly of water.

\subsubsection{Simulation set-up}
\label{section:Validation_WaterFreezing_Setup}

The speed scale was chosen as $\mathrm{U} = \nu_l/X$.
The edge length of the experimental cube's test section, taken from \cite{kowalewski1999freezing}, was used as the length scale, i.e. $\mathrm{X} = 0.038 \ \mathrm{m}$.
A realistic value for the liquid kinematic viscosity, $\nu_l = 1.0032 \times 10^{-6} \ \mathrm{m^2/s}$ as taken from \cite{michalek2003simulations}, furthermore yields the characteristic time scale $\mathrm{X}^2/\nu_l = 1440 \mathrm{s}$. 
Based on physical parameters provided in \cite{danaila2014newton} and according to the previously introduced definitions \eqref{eq:SimilarityParameters}, the similarity parameters are Ra~=~$2.52 \times 10^6$, Pr~=~6.99, and Ste~=~0.125. 
Phase-dependent thermal conductivity and volumetric heat capacity were included with $\kappa_s/\kappa_l = 3.767$ and $(\rho_s c_s)/(\rho_l c_l) = 0.4867$ in (\ref{eq:PhaseDependentConductivity}) and (\ref{eq:PhaseDependentVolumetricHeatCapacity}).
The nonlinear water density model from \cite{danaila2014newton} was used.
First proposed in \cite{gebhart1977new}, the model accounts for the density anomaly of water by expanding the density around its maximum $999.972 \ \mathrm{kg/m^{-3}}$ which occurs at $4.0293 \ \mathrm{^\circ C}$.

The simulation initial values correspond to the ``warm start'' initial conditions from \cite{kowalewski1999freezing}.
For this, the left and right walls were respectively kept constant at hot and cold temperatures $\mathpzc{T}_h = 10 \ ^\circ \mathrm{C}$ and $\mathpzc{T}_c = 0 \ ^\circ \mathrm{C}$.
A steady state convection was reached. 
For the simulation, this initial temperature range was used for the temperature scale, i.e. $\delta \mathrm{T} = \mathpzc{T}_h - \mathpzc{T}_c = 10 \ ^\circ \mathrm{C}$.
With this scale, the dimensionless boundary temperatures were $T_h = 1$ and $T_c = 0$.
The steady state problem was solved directly.
Therefore, the liquid volume fraction was set to a constant $\phi_l = 1$ and time derivatives were set to zero.
In this case, the momentum (\ref{eq:Momentum}) and energy (\ref{eq:Energy}) equations reduce to
\begin{align} 
    \label{eq:Momentum-NaturalConvection}
    \nabla\mathbf{u} \cdot \mathbf{u}
    + \nabla p - \frac{2}{\mathrm{Re}} \nabla\cdot \mathrm{sym} \nabla \mathbf{u} 
    + \frac{\mathrm{Ra}}{\mathrm{Pr}} \ b \ \mathbf{\hat{g}} 
    = 0,
    \\
    \label{eq:Energy-NaturalConvection}
    \mathbf{u} \cdot \nabla T - \frac{1}{\mathrm{Re Pr}} \nabla \cdot \left( \kappa \nabla T \right) 
    = 0
\end{align}
These equations were solved with the same mixed finite elements and Newton method as from Section \ref{section:NumericalMethods}.

\begin{figure}[h]
    \centering
    \includegraphics[width=0.45\linewidth]{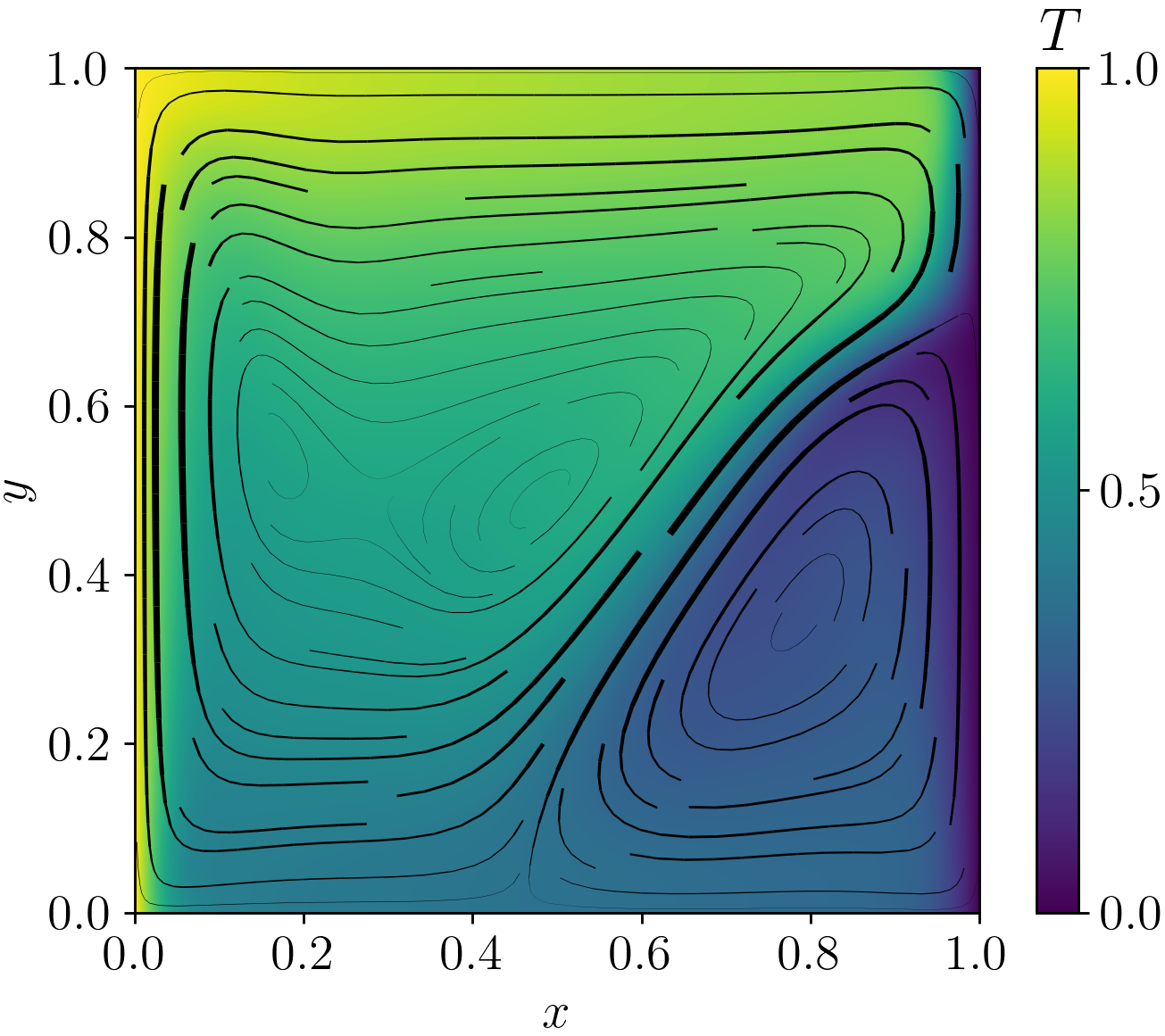}
    \caption[.]{
        Steady state solution for natural convection of water.
        
        The top and bottom walls were adiabatic. 
        The temperature of the right wall was held constant at the liquidus,
        while the temperature of the left wall was held constant at a higher temperature. 
        The temperature field is colored and the velocity streamlines are black.
        Streamline thicknesses are proportional to the local velocity magnitude.
        Two convection cells resulted from the density anomaly of water.
        }
    \label{fig:Validation_WarmStart}
\end{figure}

The high Rayleigh number caused the nonlinear solver to diverge.
Conveniently, when using the continuation module of Sapphire \cite{zimmerman2020sapphire}, which implements the procedure from Section \ref{section:Continuation}, any scalar parameter can be chosen as the continuation parameter instead of $\sigma$.
The Rayleigh number was chosen as the continuation parameter, and bounded search mode was used, bounded between values of zero and Ra.
The resulting continuation sequence was 0, Ra/16, Ra/8, Ra/4, Ra/2, and finally Ra.

To begin the time dependent simulation, the cold wall temperature was dropped to $T_c = -1$ while the hot wall temperature remained at $T_h = 1$.
This caused freezing to proceed from right to left. 
Note that the temperature scaling was unchanged.
Therefore, the physical temperature of the cold wall was $\mathpzc{T}_c = -10 \ ^\circ \mathrm{C}$, as it was in the experiment.

\subsubsection{Sensitivity to numerical parameters}

\begin{figure}[h]
    \centering
    \includegraphics[width=1.\linewidth]{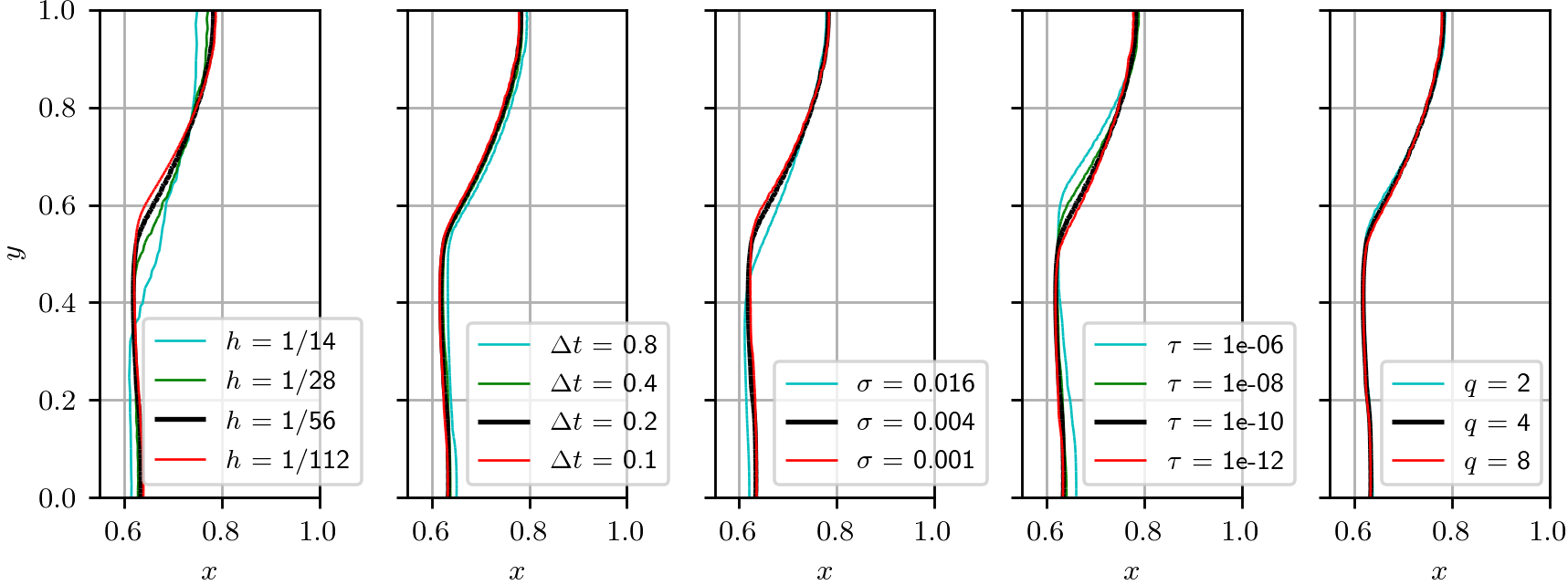}
    \caption[.]{
        Sensitivity study for the water freezing simulation.
        
        The numerical parameters in this study were the mesh cell size $h$, time step size $\Delta t$, phase interface regularization parameter $\sigma$, solid velocity relaxation parameter $\tau$, and quadrature degree $q$.
        With respect to nominal values ($h = 1/56$, $\Delta t = 0.2$, $\sigma = 0.004$, $\tau = 10^{-10}$, and $q = 4$), each parameter was varied individually and the resulting phase interfaces were compared at time $t = 1.6$.
        In each plot, the nominal result is shown in black.
        }
    \label{fig:Sensitivity_WaterFreezing}
\end{figure}

For this sensitivity study, the nominal parameter values were $h = 1/56$, $\Delta t = 0.2$, $\sigma = 0.004$, $\tau = 10^{-10}$, and $q = 4$.
From their nominal values, each parameter was varied independently.
The resulting phase interfaces, shown in Figure \ref{fig:Sensitivity_WaterFreezing}, were compared at time $t = 1.6$.
The mesh cell size, $h$, was varied from a maximum of $h = 1/14$ to a minimum of $h = 1/112$.
Using the largest size significantly degraded the accuracy.
The solution appeared to converge asymptotically as $h$ was reduced.
Between the smallest two values, there was almost no change near either the top or bottom wall, and only a minor difference in the center.
This result is better than was seen for the octadecane melting simulation in Section \ref{subsection:Validation_Melting_octadecane}, where there was a much larger sensitivity to $h$ when using the same values.

The time step size, $\Delta t$, was varied from a maximum of $\Delta t = 0.8$ to a minimum of $\Delta t = 0.1$.
Only the largest time step size significantly affected the result.
The regularization parameter, $\sigma$, was varied from a maximum of $\sigma = 0.016$ to a minimum of $\sigma = 0.001$.
The difference between the two smallest values ($\sigma = 0.002$ and $\sigma = 0.001$) was negligible.
The solid velocity relaxation parameter $\tau$ was varied from a maximum of $\tau = 10^{-6}$ to a minimum of $\tau = 10^{-12}$.
For the largest value $\tau = 10^{-6}$, significantly more freezing occurred near the center and less freezing occurred near the bottom wall.
The solution appeared to be approaching an asymptotic limit.
Between the two smallest values of $\tau$ ($\tau = 10^{-10}$ and $\tau = 10^{-12}$), the change was negligible.
The quadrature degree, $q$, was varied between $q = 2$, $q = 4$, and $q = 8$, and the effects were negligible.

\begin{figure}[h]
    \centering
    \includegraphics[width=1.\linewidth]{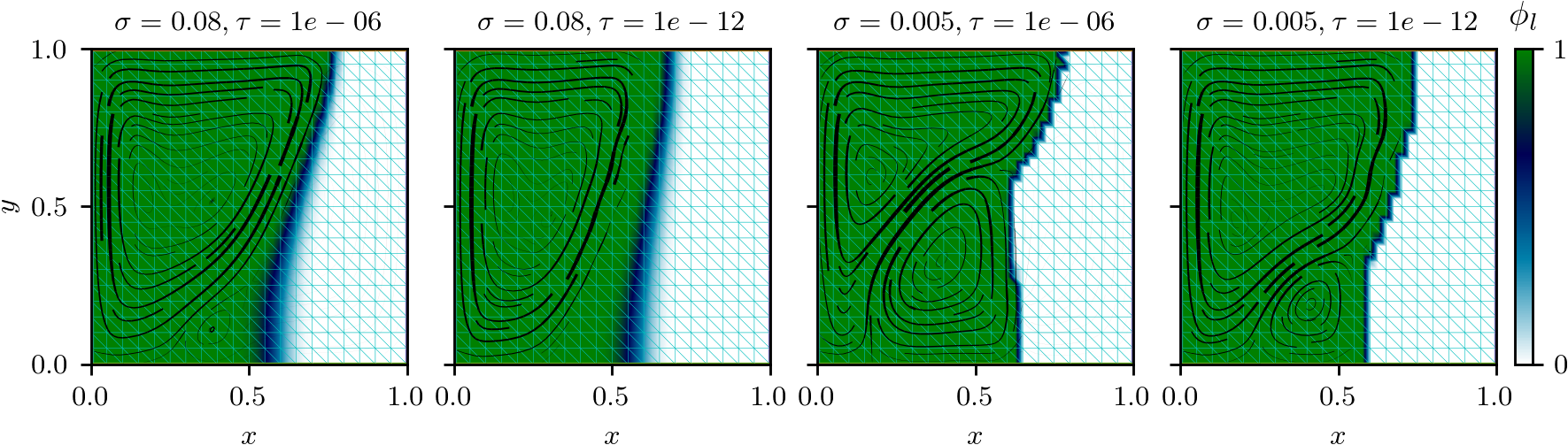}
    \vskip 0.5\baselineskip
    \includegraphics[width=1.\linewidth]{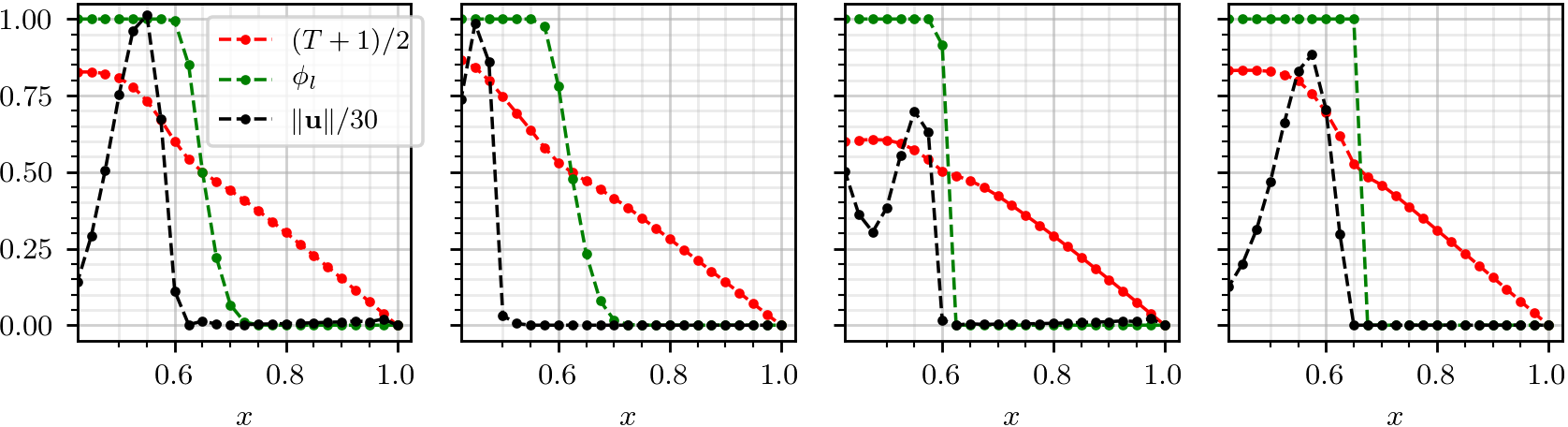}
    \vspace{-8mm}
    \caption[.]{Example 1D profiles from water freezing solution.
    
        The simulation set-up is the same as the water freezing benchmark in Section \ref{section:Validation_WaterFreezing_Setup}.
        Four solutions are shown at $t = 2$ with varied $\sigma$ and $\tau$.
        Other numerical parameters were $h = 0.05$, $\Delta t = 1$, and $q= 4$.
        Top) The mesh in cyan, velocity streamlines in black, and colored $\phi_l$.
        Bottom) 1D solution profiles at $y = 0.5$. $T$ and $\mathbf{u}$ are normalized.
        To focus on the solid region, the $x$ axis spans only the right half of the domain.
        }
    \label{fig:WaterFreezing-VarySigmaTau}
\end{figure}

Figure \ref{fig:WaterFreezing-VarySigmaTau} further demonstrates the effects of varying $\sigma$ and $\tau$ on a constant coarse mesh.
The size of the artificial mushy region (where $0 < \phi_l < 1$) varied with $\sigma$.
For the case with largest $\sigma$ and smallest $\tau$, the velocity solution was disturbed so greatly that the convection-cell in the lower-right of the liquid region disappeared entirely.
The one-dimensional profiles on the bottom of Figure \ref{fig:WaterFreezing-VarySigmaTau} further demonstrate the effect of the velocity relaxation in the artificial mushy region.
For both cases with $\sigma = 0.005$, the artificial mushy region was contained by a single cell (though for the $\tau = 10^{-6}$ case, a point with $0 < \phi_l < 1$ is visible, because there are two temperature degrees of freedom in each direction of each cell).
With $\sigma = 0.08$, the artificial mushy region extended over multiple cells.
For both cases with $\tau = 10^{-12}$, the velocity was approximately zero at the left edge of the artificial mushy region and remained zero throughout the solid.
For both cases with $\tau = 10^{-6}$, the velocity was substantially above zero at the left edge of the artificial mushy region (though for the $\sigma = 0.005$ case, this is harder to see). 
Furthermore, for the $\tau = 10^{-6}$ cases, while the velocity reached approximately zero at the left edge of the solid region, it steadily increased while moving further into the solid region, until the rightmost cell where it again dropped to zero to meet the boundary condition.

\subsubsection{Results and discussion}
\label{section:Validation_WaterFreezing_Results}
\begin{figure}[h]
    \includegraphics[width=1.\linewidth]{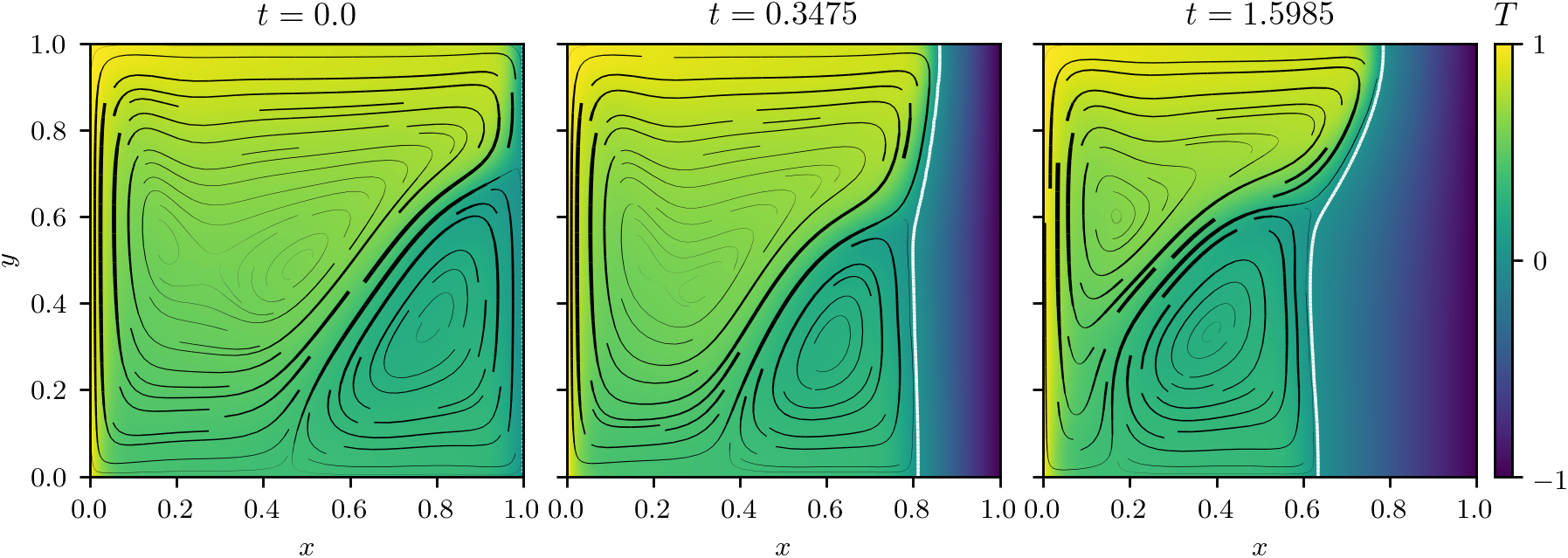}
    \caption[.]{
        Water freezing simulation results at three times.
        
        Two convection cells are visible in the liquid domain.
        Freezing proceeded from the right wall toward the left.
        The temperature field is colored and the velocity streamlines are black with thickness proportional to the velocity magnitude.
        Left) The initial steady state solution before dropping the right wall's temperature. 
        Center and Right) A white line marks the liquidus isotherm.
        }
    \label{fig:Validation_WaterFreezing_Simulation}
\end{figure}

The sensitivity study revealed that large time step sizes $\Delta t$ can be used without significantly changing the resulting phase interface.
To most easily compare to experimental results in \cite{kowalewski1999freezing}, solutions were obtained at every one-hundred physical seconds, corresponding to a simulated time step size of about $\Delta t =  0.0695$.
The other numerical parameters were $h = 1/112$, $\sigma~=~0.004$, $\tau = 10^{-10}$, and $q = 4$.
Figure \ref{fig:Validation_WaterFreezing_Simulation} visualizes results at the initial steady state and two later times, $t = 0.3475$ (500 seconds) and $t = 1.5985$ (2300 seconds).
The temperature field is colored and the velocity streamlines are black with thickness proportional to the velocity magnitude.
As freezing proceeded from right to left, the two circulating regions of natural convection were maintained, translating to the left along with the phase interface.
The bottom circulating region showed a lower temperature and velocity. The freezing front was nearly planar and proceeded more rapidly in this region.
The top circulating region had higher temperature and velocity. The freezing front was curved and proceeded less rapidly near the top wall.
Also, near the phase interface, note the sharper temperature gradient near the top wall.
This slowed the freezing process in that region.

By the final time, after a total of twenty-three time steps, 1910 Newton iterations were used in total, i.e. about 83 Newton iterations per time step.
This includes all iterations from intermediate problems during the continuation procedure sketched in Figure \ref{fig:continuation}.
For the first time step, eight intermediate values of $\sigma$ = 0.512, 0.258, 0.131, 0.0675, 0.03575, 0.0278125, 0.019875, and 0.0119375 were used for continuation.
At $t = 0.417$ an additional intermediate value of $\sigma = 0.09925$ was used, which sufficed for the rest of the simulation.
As shown in Figure \ref{fig:continuation}, each step in continuation corresponds to solving a nonlinear system with Newton's method.
For any given time step in this simulation, the average number of Newton iterations per continuation step was between eight and nine.

\begin{figure}[h]
    \includegraphics[width=1.\linewidth]{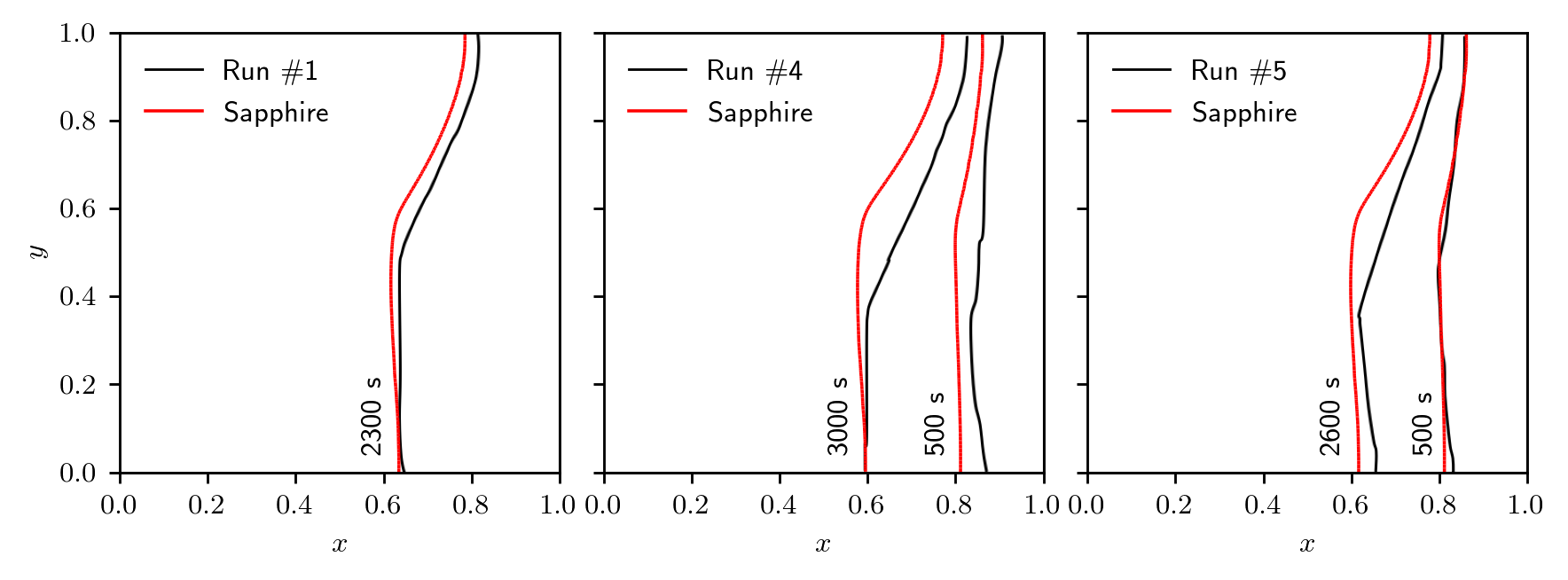}
    \caption[.]{
        Freezing simulation results compared to multiple experimental runs from \cite{kowalewski1999freezing}. 
        
        The same simulation result is compared to three experimental runs which differed in their initial states.
        Run \#1 used ``cold start'' initial conditions described in \cite{kowalewski1999freezing}.
        Runs \#4 and \#5 used ``warm start'' initial conditions briefly described in Section \ref{section:Validation_WaterFreezing_Results} and detailed in \cite{kowalewski1999freezing}.
        The simulation initial values were based on the ``warm start'' procedure, as described in Section \ref{section:Validation_WaterFreezing_Results}.
        The experimental runs also differed regarding the thermal control used at the walls of the test section.
        Runs \#1 and \#4 had a constant air flow outside the walls.
        For run \#5, the test section was instead submerged in a water bath.
        }
    \label{fig:Validation_WaterFreezing_CompareToExperiment}
\end{figure}

Figure \ref{fig:Validation_WaterFreezing_CompareToExperiment} compares the simulation results to a group of experimental runs from \cite{kowalewski1999freezing}.
The simulation compares remarkably well to experimental run \#1; but that run reportedly used a ``cold start''.
The simulation instead used a ``warm start'', which corresponds to the experimental runs \#4 and \#5.
It was also shown in \cite{kowalewski1999freezing} that the freezing process was sensitive to the thermal control of the test section walls.
Experimental runs \#1 and \#4 had a constant air flow around the walls.
For run \#5, the test section was submerged in a water bath.
The importance of simulating three-dimensional heat transfer in the side walls was further demonstrated in \cite{giangi2000natural}. 
There, it was also noted that the viscosity of water substantially increases near its freezing temperature, and that this in return substantially affects the freezing process.
Further investigating the effects of a temperature dependent viscosity or heat transfer through the test section walls was outside the scope of this work.

\begin{figure}[h]
    \centering
    \includegraphics[width=0.8\linewidth]{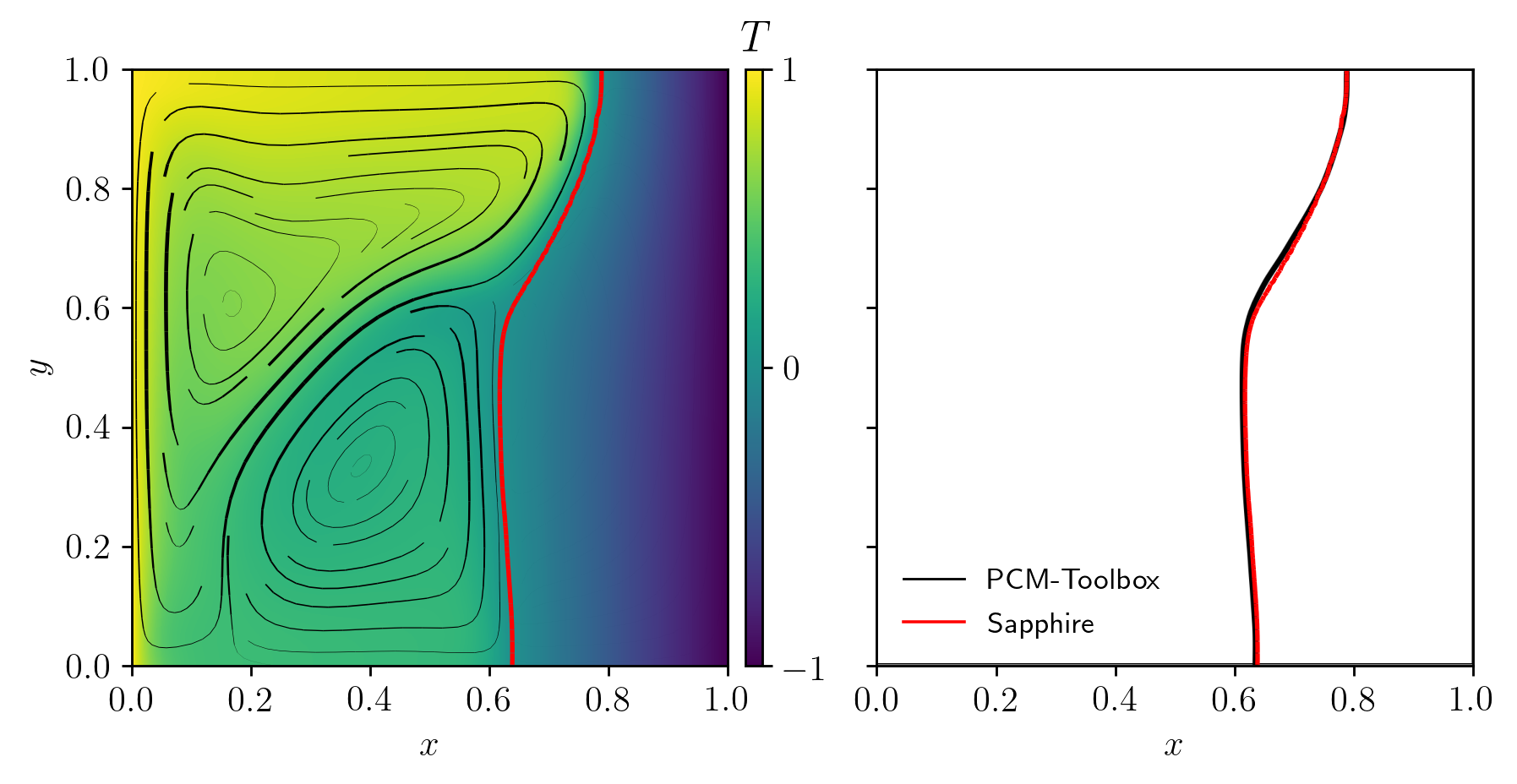}
    \caption[.]{
        Comparison to water freezing result in \cite{rakotondrandisa2020toolbox} at time $t = 1.6$.
        
        The Sapphire simulation's numerical parameters were $h = 0.02$, $\Delta t = 0.1$, $\sigma = 0.00875$, $\tau = 10^{-7}$, and $q = 4$.
        }
    \label{fig:Validation_WaterFreezing_CompareToPCMToolbox}
\end{figure}

Figure \ref{fig:Validation_WaterFreezing_CompareToPCMToolbox} compares a simulation using Sapphire to results presented in \cite{rakotondrandisa2020toolbox}.
As when comparing to their octadecane melting simulation,
the regularization parameter was set to $\sigma = 0.00875$ 
and simulations were ran with $10^{-6} < \tau < 10^{-8}$.
In the simulation from \cite{rakotondrandisa2020toolbox}, the mesh was adaptively refined at each time step, resulting in meshes of less than 3000 vertices.
That simulation required several days \cite{rakotondrandisa2020toolbox} of compute time.
Sapphire's simulation, using 2500 uniformly distributed mesh vertices (i.e. $h = 0.02$), required fifteen minutes of compute time (having solved a total of 784 Newton iterations) and yielded a nearly identical result.

\section{Additional examples}
\label{section:Examples}

Two additional examples demonstrate the flexibility of the method and code: a higher Rayleigh number octadecane melting example based on \cite{bertrand1999melting} and a gallium (lower Prandtl number) melting example based on \cite{belhamadia2019adaptive}.
Both simulation setups were mostly similar to the octadecane melting simulation in Section \ref{subsection:Validation_Melting_octadecane}.
The same code was run on the same CPU as in Section \ref{subsection:Validation_Melting_octadecane}.

\subsection{Melting octadecane at a higher Rayleigh number}

This example is characterized primarily by the higher Rayleigh number of Ra = $10^7$, where Ra = 3.27 $\times 10^5$ was used in Section \ref{subsection:Validation_Melting_octadecane}.
The other similarity parameters were Pr = 50 and Ste = 0.1.
The length scale was X = 0.1 meters and the time scale was $\mathrm{X}^2/\nu_l = 1,000$ seconds.
A rectangular domain was used with $L_x = 0.5$.
The numerical parameters were $h = 1/80$, $\Delta t = 0.05$, $\sigma = 0.005$, $\tau = 10^{-10}$, and $q = 4$.

\begin{figure}[h]
    \centering
    \includegraphics[width=0.9\linewidth]{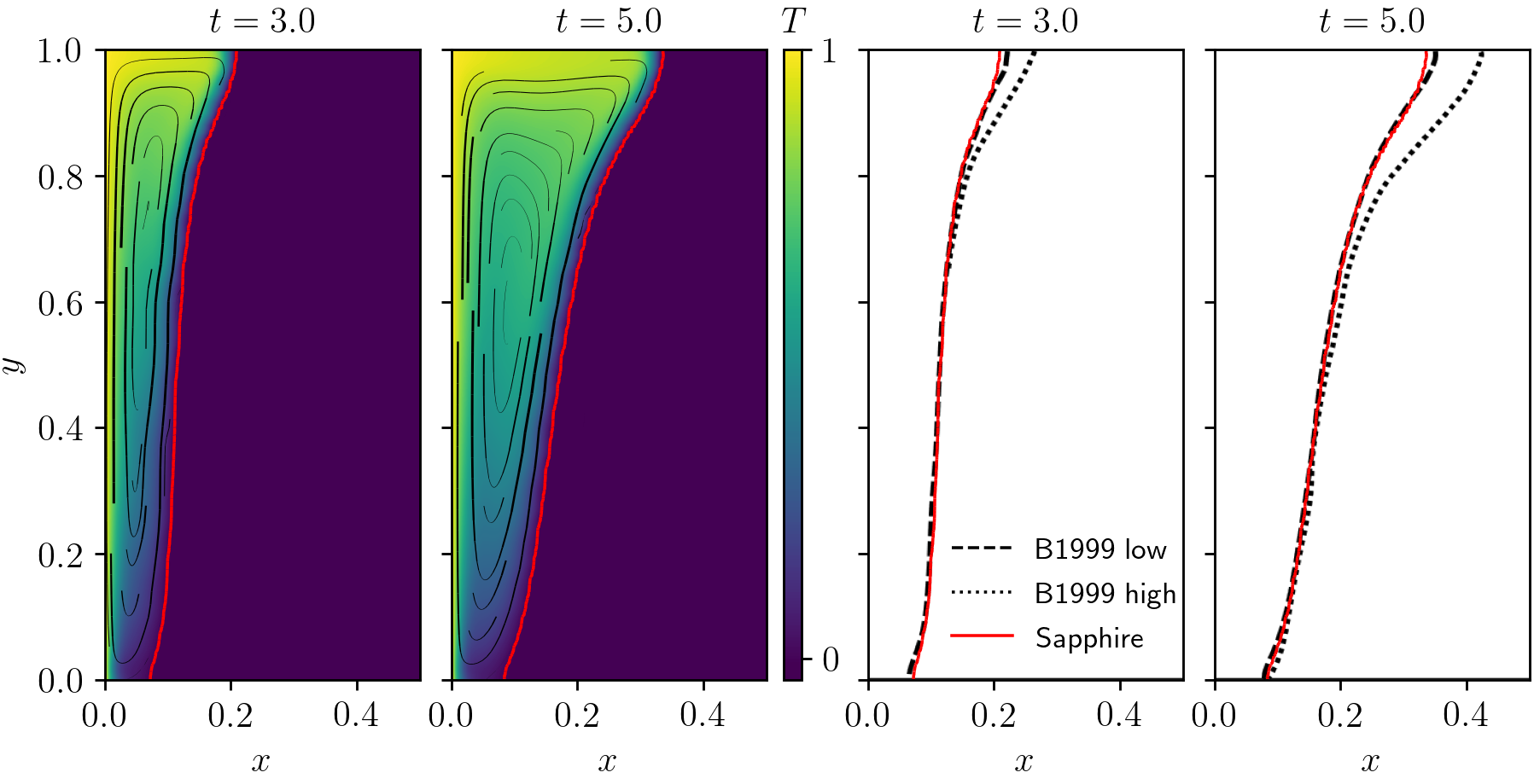}
    \caption[.]{
        Higher Rayleigh number octadecane melting simulation and comparison to \cite{bertrand1999melting}.
        
        Left) Temperature field (colored), phase interface (red) and velocity streamlines (black).
        Right) Phase interface overlaid with range of accepted results in \cite{bertrand1999melting}.
        }
    \label{fig:HighRayleigh}
\end{figure}

By the final time, i.e. after one-hundred time steps, a total of 2900 Newton iterations were used including all continuation steps.
Up to four intermediate regularizations were required for continuation at each time step.
The simulation required one hour and three minutes of compute time.
The result is shown in Figure \ref{fig:HighRayleigh} and nearly matches the accepted result from \cite{bertrand1999melting} that predicted the least melting.

\subsection{Melting gallium}

This example is characterized primarily by a smaller Prandtl number of Pr = 0.0216.
The other similarity parameters were Ra = $7 \times 10^5$ and Ste = 0.046.
Unlike the other simulations presented in this work, the velocity scale was chosen as $\mathrm{U} = \alpha_l/\mathrm{X}$ and therefore the Reynolds number appearing in the dimensionless governing equations was Re = 1/Pr.
The nondimensional temperature initial value and cold wall boundary condition was $T_c = -0.1546$.
The length scale was X = 0.0635 meters and the time scale was $\mathrm{X}^2/\alpha_l = 292.9$ seconds.
A rectangular domain was used with $L_x = 0.25$.
Using $h = 1/160$, and again a uniform mesh, there were a total of 12,800 mesh cells.
The other numerical parameters were $\Delta t = 0.001$, $\sigma = 0.01$, $\tau = 10^{-10}$, and $q = 4$.

\begin{figure}[h]
    \centering
    \hspace{0.3cm}
    \includegraphics[width=0.85\linewidth]{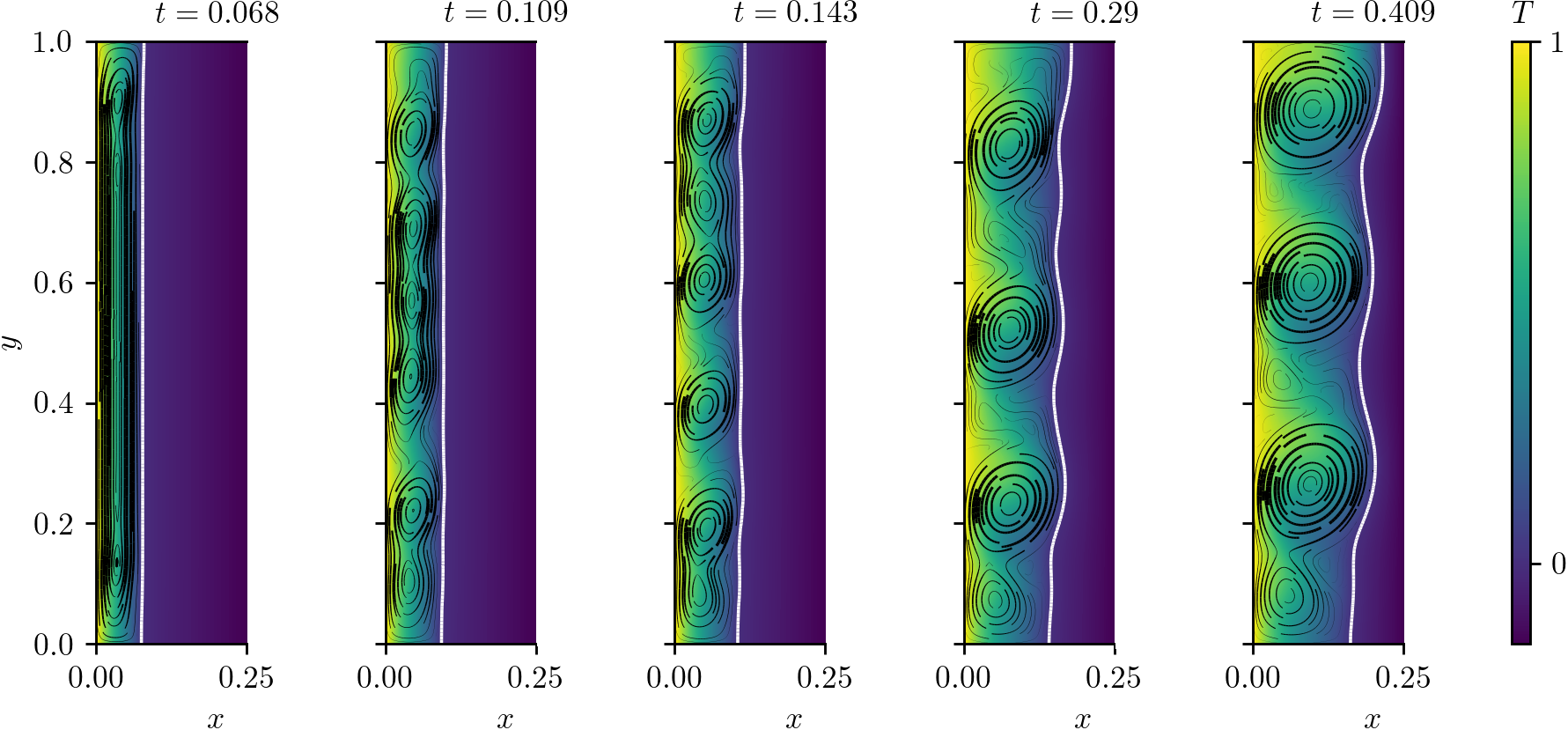}
    \includegraphics[width=0.82\linewidth]{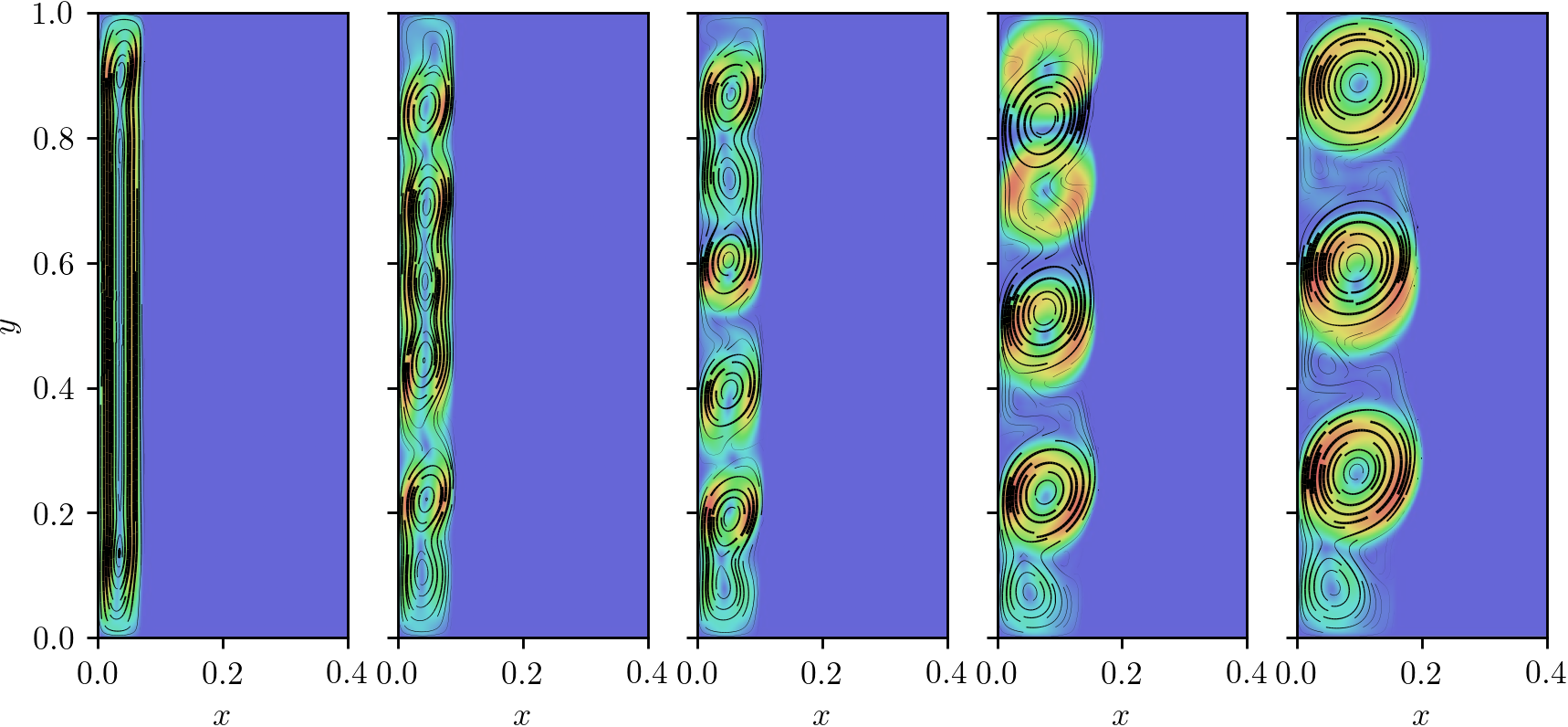}
    \vspace{-0.2cm}
    \caption[.]{
        Gallium melting simulation and comparison to \cite{belhamadia2019adaptive}.
        
        Top) Temperature field, black velocity streamlines, and white liquidus isotherm from Sapphire simulation shown at five times for comparison to \cite{belhamadia2019adaptive}.
        Bottom) Black velocity streamlines from Sapphire simulation overlaid with velocity magnitude field (blue is zero, red is maximum) from \cite{belhamadia2019adaptive}.
        }
    \label{fig:GalliumMelting}
\end{figure}

By the final time, i.e. after 409 time steps, a total of 5700 Newton iterations were used including all continuation steps.
Up to five intermediate regularizations were required for continuation at each time step.
The simulation required three hours and fifty minutes of compute time.
Figure \ref{fig:GalliumMelting} shows the result.
Compared to the results from \cite{belhamadia2019adaptive}, where adapted meshes with an average of 12,000 mesh cells were used on a rectangular domain with $L_x = 0.5$,
the largest discrepancy is with the upper convection cells at time $t = 0.29$.
An additional convection cell developed in the \cite{belhamadia2019adaptive} result.
Overall, the results compare remarkably well.

\section{Conclusions}
\label{section:Conclusions}

An enthalpy method was used to simulate convection-coupled isothermal phase-change on a single geometric domain.
The governing equations were discretized in space with mixed finite elements and in time with backward difference formulas.
The resulting system of nonlinear equations was solved with Newton's method.
As new contributions, the current work 
\begin{itemize}
    \itemsep0em
    \item presented a new continuation procedure that reliably converged Newton's method for the tested 2D isothermal phase-change benchmarks.
    Without this procedure, previous literature such as \cite{danaila2014newton} and \cite{ zimmerman2017monolithic} relied on carefully selected time step sizes, specifying a one-way melting or solidification process, initializing the new phase in part of the domain, multiple approaches to adaptive mesh refinement, and manually refining the initial mesh, in order to achieve convergence of Newton's method.
    The key idea of the new continuation procedure is to recognize that the regularity of the nonlinear problem's solution space is dominated by a global scalar parameter, the regularization parameter $\sigma$.
    The procedure highlights a path forward for regularizing this class of nonlinear problems.
    \item shared the open source code in a new Python packaged called Sapphire \cite{zimmerman2020sapphire},
    using the finite element library Firedrake \cite{rathgeber2016firedrake}. 
    Sapphire's test suite covers many of this paper's results. These tests can be used as scripts to reproduce the results.
    The tests are separated into verification tests and validation tests.
    The validation tests include both benchmark problems from this paper, using coarse meshes and large time steps so that they run quickly.
    The versions of Sapphire and Firedrake were documented with DOIs on Zenodo, respectively at \cite{zimmerman2020sapphire} and \cite{zenodo/Firedrake-20200611.3}.
    \item verified the implementation via a formal convergence study using the method of manufactured solutions. 
    Second order spatial accuracy was verified in 2D for the velocity and temperature. 
    The pressure's spatial discretization error showed super-convergence, which was not further explored in this contribution.
    Second order temporal accuracy was verified for all solution components which were discretized in time, i.e. the velocity and the temperature.
    \item presented benchmark simulations for melting octadecane and freezing water in square cavities. 
    For each, sensitivities were studied with respect to the mesh cell size $h$, time step size $\Delta t$, regularization parameter $\sigma$, solid velocity relaxation parameter $\tau$, and quadrature degree $q$.
    For the freezing water case, results were obtained which were not largely sensitive to the five numerical parameters $h$, $\Delta t$, $\sigma$, $\tau$, or $q$.
    For the octadecane melting case, the refined result was still largely sensitive to $h$.
    Further reducing $h$ was impractical, because it would increase the number of degrees of freedom in the linear system quadratically, which is particularly prohibitive for the direct linear system solver.
    Reducing $h$ could affect sensitivity in the other parameters, especially $\sigma$.
    \item compared the benchmark simulation results to experimental data sets from the literature, namely \cite{okada1984analysis}  for melting octadecane and \cite{kowalewski1999freezing} for freezing water.
    For the melting octadecane case, the simulation under-predicts melting with respect to the experiment. The largest discrepancy is near the top wall at the end of the simulation.
    Reducing the mesh cell size $h$ should reduce the discrepancy near the top wall, but no such improvement should be expected near the bottom wall or interior.
    For the water freezing case, multiple experimental runs were considered.
    The runs varied with respect to initial conditions and thermal regulation of the test section walls.
    While the simulation compared well to one of the runs, that run used a different procedure for initialization which is not considered by the simulation.
    More information is needed to accurately define the initial conditions for the water freezing case, and adiabatic wall assumptions are too ideal for both cases.
    \item compared the benchmark simulation results to those from \cite{rakotondrandisa2020toolbox} which used adaptive mesh refinement.
    The results compare favorably, even when using uniform meshes with a similar number of cells.
    \item presented two additional example simulations, one for melting octadecane with a higher Rayleigh number, and another for melting gallium.
    In both cases, results compare favorably with those in the literature.
\end{itemize}

The application scope of the Sapphire software \cite{zimmerman2020sapphire} is potentially much larger than demonstrated in the current work.
Having used the finite element method, application to more complicated geometries should be straightforward.
By using UFL \cite{alnaes2012unified} for symbolically defining the mathematical model and Firedrake \cite{rathgeber2016firedrake} for automating most of the implementation, the model and implementation are easily modifiable and extensible.
The continuation procedure has been an effective tool for reliably solving the nonlinear problems in this work for a large range of parameters.
Still, there should exist a more general, less ad hoc, approach to regularizing this class of nonlinear problems. 
Finding such an approach could further decrease computational costs and further improve robustness.
There are other promising routes for reducing computational costs. 
One route would be the application of iterative linear system solvers, which would require the development of a preconditioner.
Implementing the iterative method and preconditioner should be straightforward using Firedrake.
Successfully applying an iterative solver would make it practical to solve problems with three-dimensional geometries, and would also open opportunities for addressing inverse problems.

\section*{Acknowledgments}
The authors were funded in part by the Excellence Initiative of the German Federal and State Governments through grant GSC 111.
The work was furthermore supported by the Federal Ministry of Economic Affairs and Energy, on the basis of a decision by the German Bundestag (50 NA 1502).





\bibliographystyle{references_style}
\bibliography{main}







\end{document}